\documentclass[dvipdfmx,aps,prd,showpacs,preprintnumbers,superscriptaddress,nofootinbib,twocolumn]{revtex4}%
\usepackage[dvipdfmx]{graphicx}
\usepackage{bm,latexsym,amsmath,amssymb,amsfonts,mathrsfs,subfigure}
\usepackage{color}
\input{colordvi.tex}
\usepackage[dvipdfmx]{hyperref}
\usepackage{url}
\hypersetup{
    colorlinks=true,
    citecolor=cyan,
}
\allowdisplaybreaks[1]
\newcommand*{\D}{{\rm d}}
\newcommand*{\mpl}{M_{\rm Pl}}

\begin{document}

\title{Primordial non-Gaussianity from Galilean Genesis without strong coupling problem}

\author{Shingo~Akama}
\email[Email: ]{shingo.akama"at"uj.edu.pl}
\affiliation{Faculty of Physics, Astronomy and Applied Computer Science, Jagiellonian University, 30-348 Krakow, Poland
}
\author{Shin'ichi~Hirano}
\email[Email: ]{hirano.s.ai"at"m.titech.ac.jp}
\affiliation{Department of Physics, Tokyo Institute of Technology, 2-12-1 Ookayama, Meguro-ku, Tokyo 152-8551, Japan
}

\begin{abstract}
Galilean Genesis is generically plagued with a strong coupling problem, but this can be avoided depending on the hierarchy between a classical energy scale of genesis and a strong coupling scale. In this paper, we investigate whether or not the models of Galilean Genesis without the strong coupling problem can explain the statistical properties of the observed CMB fluctuations based on two unified frameworks of Galilean Genesis. By focusing on the class in which the propagation speeds of the scalar and tensor perturbations are constant, we show that the models avoiding strong coupling and allowing a slightly red-tilted scalar power spectrum suffer from an overproduction of a scalar non-Gaussianity.

\end{abstract}

\maketitle

\section{Introduction}

Inflation~\cite{Guth:1980zm,Starobinsky:1980te,Sato:1980yn} is currently the standard paradigm of the early universe, even though it suffers from the initial singularity~\cite{Borde:1996pt}. To avoid the singularity, non-singular alternative paradigms with the violation of the null energy condition have also been studied so far (see e.g., Refs.~\cite{Creminelli:2010ba} and~\cite{ Battefeld:2014uga}). Inflation not only superbly resolves various problems in the standard Big-Bang cosmology but also successfully explains the origins of both the CMB anisotropies and the rich structure of our universe. Of particular interest is to see whether such non-singular paradigms can truly be alternatives to inflation or not from both theoretical and observational perspectives.

Galilean Genesis~\cite{Creminelli:2010ba} is one of the non-singular alternative scenarios in which the universe is quasi-Minkowski in asymptotic past. This scenario can resolve the problems in the standard Big-Bang cosmology as well as inflation~\cite{Nishi:2015pta,Nishi:2016ljg}. As the generic theoretical problem, the non-singular scenarios have been found to be plagued with the occurrence of gradient instabilities in scalar perturbations~\cite{Libanov:2016kfc,Kobayashi:2016xpl,Cai:2016thi,Creminelli:2016zwa,Akama:2017jsa}, though how to overcome that has been clarified (see e.g., Refs.~\cite{Cai:2016thi,Creminelli:2016zwa,Cai:2017tku,Cai:2017dyi,Kolevatov:2017voe,Boruah:2018pvq,Heisenberg:2018wye,Ye:2019frg,Ye:2019sth,Quintin:2019orx}). Besides this problem, Galilean Genesis suffers from another theoretical problem: strong coupling at an early stage of genesis~\cite{Ageeva:2018lko,Ageeva:2020gti,Ageeva:2020buc,Ageeva:2021yik}, which indicates that one can trust neither any analyses based on the perturbation theory at the early stage nor any successful observational predictions consistent with CMB data based on it. One can avoid this problem as long as some typical energy scale of genesis is much lower than the scale at which strong coupling occurs~\cite{Ageeva:2018lko,Ageeva:2020gti,Ageeva:2020buc,Ageeva:2021yik,Ageeva:2022fyq,Ageeva:2022asq}.

So far the primordial power spectra of the scalar and tensor perturbations have been well studied based on unified frameworks of Galilean Genesis~\cite{Nishi:2015pta,Nishi:2016ljg} proposed in the Horndeski theory, the most general single-scalar-tensor theory with second-order field equations~\cite{Horndeski:1974wa,Deffayet:2011gz,Kobayashi:2011nu} (see also Ref.~\cite{Kobayashi:2019hrl} for a review). Whereas the strong coupling problem has been studied in subclasses of the unified frameworks~\cite{Ageeva:2018lko,Ageeva:2020gti,Ageeva:2020buc,Ageeva:2021yik,Cai:2022ori} (and in an example outside  the frameworks~\cite{Cai:2022ori}). 
Also, the primordial non-Gaussianities are important quantities as well from both theoretical and observational viewpoints, though those have not been calculated in the context of Galilean Genesis so far. The main purpose of the present paper is to see whether or not Galilean Genesis without the theoretical problems can predict the observational signatures consistently with the observed CMB fluctuations. In the present paper, 
after revisiting the strong coupling problem in the unified frameworks and clarifying the parameter region of the models predicting the slightly red-tilted scalar power spectrum, we evaluate the scalar non-Gaussianity in the allowed parameter region and discuss whether the models can enjoy all of the observational constraints on the early universe models.

This paper is organized as follows. In the following section, we introduce two unified frameworks of Galilean Genesis. In Sec.~\ref{sec: power}, we give a brief review of the primordial power spectra for scalar and tensor perturbations in both frameworks. In Sec.~\ref{sec: strong coupling}, we first make arguments on the strong coupling problem and then clarify the model space of Galilean Genesis with the scale-invariant scalar power spectrum and without strong coupling. In Sec.~\ref{sec: non-G}, we calculate the scalar non-Gaussianity in the model space and compare the non-Gaussianity of the curvature perturbation with the current constraints on that. A summary of our paper is given in Sec.~\ref{sec: summary}.

\section{Frameworks}
In the present paper, we assume a spatially flat Friedmann-Lema\^{i}tre-Robertson-Walker (FLRW) metric of the form
\begin{align}
\D s^2=-\D t^2+a^2(t)\delta_{ij}\D x^i\D x^j,
\end{align}
where $a(t)$ is the scale factor.

So far most models of Galilean Genesis have been studied within the Horndeski theory~\cite{Creminelli:2010ba,Piao:2010bi, Liu:2011ns,PerreaultLevasseur:2011wto,Wang:2012bq,Creminelli:2012my,Hinterbichler:2012fr,Hinterbichler:2012yn,Easson:2013bda,Rubakov:2013kaa,Elder:2013gya,Cai:2016gjd}. (See Refs.~\cite{Kobayashi:2015gga,Cai:2017tku,Mironov:2019qjt, Mironov:2019haz,Volkova:2019jlj,Ilyas:2020zcb,Zhu:2021ggm,Cai:2022ori} for the models studied in beyond Horndeski theories~\cite{Gleyzes:2014dya,Langlois:2015skt,BenAchour:2016fzp,Langlois:2017mxy,Kobayashi:2019hrl}.) The Horndeski theory yields the most general second-order field equations for a scalar field $\phi$ and a metric $g_{\mu\nu}$, and the Lagrangian of which is~\cite{Horndeski:1974wa,Deffayet:2011gz,Kobayashi:2011nu}
\begin{align}
\mathcal{L}&=G_2(\phi,X)-G_3(\phi,X)\Box\phi
+G_4(\phi,X)R
\notag \\ &\quad
+G_{4X}\left[(\Box\phi)^2-(\nabla_{\mu}\phi\nabla_{\nu}\phi)^2\right]
\notag \\ &\quad
+G_5(\phi,X)G_{\mu\nu}\nabla^{\mu}\nabla^{\nu}\phi
-\frac{G_{5X}}{6}\bigl[(\Box\phi)^3
\notag \\ &\quad
-3\Box\phi(\nabla_{\mu}\nabla_{\nu}\phi)^2+2(\nabla_{\mu}\nabla_{\nu}\phi)^3\bigr],\label{eq: Horndeski action}
\end{align}
where $G_{i}~(i=2,\cdots,5)$ are arbitrary functions of $\phi$ and $X$ with $X:=-g^{\mu\nu}\nabla_\mu\phi\nabla_\nu\phi/2$ being the kinetic term of the scalar field, and $G_X$ stands for the partial derivative of $G$ with respect to $X$, i.e., $G_X=\partial G/\partial X$. From the Horndeski action, one can derive the Friedmann and evolution equations denoted by $\mathcal{E}=0$ and $\mathcal{P}=0$, respectively, as written in Appendix.~\ref{sec: background}.


In the scenario of Galilean Genesis~\cite{Creminelli:2010ba}, the cosmic expansion starts from a quasi-Minkowski phase. In particular, by assuming the following configuration of the scalar field
\begin{align}
Y:= e^{-2\lambda\phi}X\simeq Y_0={\rm const.}, \label{ansatz-phi}
\end{align}
giving
\begin{align}
e^{\lambda\phi}&\simeq\frac{1}{\lambda\sqrt{2Y_0}}\frac{1}{(-t)}, \label{sol-phi}
\end{align}
the Lagrangian with appropriate choices of $G_i$ admits a quasi-Minkowski solution where the Hubble parameter $H:=(\D a/\D t)/a$ is asymptotic to $0$ by a power-law manner in asymptotic past (a large $|t|$ region). 
In the following subsection, we briefly review two unified frameworks of Galilean Genesis.

\subsection{Generalized Galilean Genesis}
The model of Galilean Genesis was originally proposed in Ref.~\cite{Creminelli:2010ba}, and each $G_i$ of the original model is of the form
\begin{align}
G_2&=c_1e^{2\lambda\phi}X+c_2 X^2,\ G_3=c_3X,\notag\\
G_4&=\frac{\mpl^2}{2},\ G_5=0,
\end{align}
with constant $c_i$.
The generalization in a way to include the original model has been accomplished by choosing the Horndeski functions as~\cite{Nishi:2015pta}
\begin{align}
G_2&=e^{2(\alpha+1)\lambda\phi}g_2(Y),\ G_3=e^{2\alpha\lambda\phi}g_3(Y),\notag\\
G_4&=\frac{\mpl^2}{2}+e^{2\alpha\lambda\phi}g_4(Y),\ G_5=e^{-2\lambda\phi}g_5(Y),
\end{align}
where $g_i(Y)$ $(i=2,\cdots,5)$ are arbitrary functions of $Y$, and $\alpha$ and $\lambda$ are constant. Note that $g_2=c_1Y+c_2Y^2,\ g_3=c_3Y,\ g_4=g_5=0$, and $\alpha=1$ for the original model. This unified framework is called Generalized Galilean Genesis. Starting with the discovery of the original model, various models included in this framework have been constructed in Ref.~\cite{Piao:2010bi, Liu:2011ns,PerreaultLevasseur:2011wto,Wang:2012bq,Creminelli:2012my,Hinterbichler:2012fr,Hinterbichler:2012yn,Easson:2013bda,Rubakov:2013kaa,Elder:2013gya}. 

By assuming Eq.~(\ref{ansatz-phi}), the quasi-Minkowski solution has been obtained in Generalized Galilean Genesis as~\cite{Nishi:2015pta}
\begin{align}
a&\simeq1+\frac{1}{2\alpha}\frac{h_0}{(-t)^{2\alpha}}=1+\mathcal{O}(Ht),
\end{align}
where 
\begin{align}
H&\simeq\frac{h_0}{(-t)^{1+2\alpha}},\ \label{eq: GGG-H}\\
\alpha&>0 \label{eq: GGG-a}.
\end{align}
The background spacetime remains the quasi-Minkowski one as long as 
\begin{align}
H|t|\ll1. \label{smallHt}
\end{align}
In this framework, the Friedmann and evolution equations have been found, respectively, to be~\cite{Nishi:2015pta}
\begin{align}
\mathcal{E}&\simeq e^{2(1+\alpha)\lambda\phi}\hat\rho_1(Y_0)=0 \label{eq: GGG-Fried},\\
\mathcal{P}&\simeq 2\mathcal{G}_1\dot H+e^{2(1+\alpha)\lambda\phi}\hat p_1(Y_0)=0, \label{eq: GGG-evo}
\end{align}
where 
\begin{align}
{\cal G}_1 &:=\mpl^2-4\lambda Y_0(g_5+Y_0 g_5'),~
\\
\hat\rho_1(Y)&:=2Yg_2'-g_2-4\lambda Y(\alpha g_3-Y g_3'),\label{rho1}\\
\hat p_1(Y)&:=g_2-4\alpha\lambda Yg_3\notag\\
&\quad\ +8(2\alpha+1)\lambda^2Y(\alpha g_4-Yg_4'), \label{p1}
\end{align}
and the dot and prime denote differentiations with respect to $t$ and $Y$, respectively. Eqs.~(\ref{eq: GGG-Fried}) and~(\ref{eq: GGG-evo}) are used to determine the values of $Y_0$ and $h_0$, respectively. In particular, the evolution equation is of the linear equation for $h_0$, and $h_0$ can easily be obtained and written as a compact form~\cite{Nishi:2015pta},
\begin{align}
h_0=-\frac{1}{2(1+2\alpha)(2\lambda^2 Y_0)^{1+\alpha}}\frac{\hat p_1(Y_0)}{\mathcal{G}_1}. \label{h0}
\end{align}
The various aspects of the background dynamics have been also investigated in the presence of the spatial curvature and spacetime anisotropies~\cite{Nishi:2015pta}. In Ref.~\cite{Nishi:2015pta}, it has been shown that both do not spoil the background evolution under Eq.~(\ref{eq: GGG-a}).

\subsection{New Framework}
In the Horndeski theory, a model having the quasi-Minkowski solution has been studied outside Generalized Galilean Genesis as well~\cite{Cai:2016gjd}. That model is based on the Lagrangian with $G_i$ having the following $\phi$- and $X$-dependence\footnote{The explicit form of $G_4$ in the Lagrangian in Ref.~\cite{Cai:2016gjd} is $G_4=(\mpl^2/2)(1+\lambda^8/X^2)$. The first term in the bracket of this $G_4$ is negligible at some large $|t|$, compared to the second one since $X^{-2}\propto(-t)^{4}$. By using this fact, we ignored the first term in Eq.~(\ref{ex.new}).}:
\begin{align}
G_2&=-d\lambda^4 e^{6\lambda\phi}-e^{4\lambda\phi}X+\lambda^8 X^3,\ G_3=0,\notag\\
G_4&=\frac{\mpl^2\lambda^8}{2X^2},\ G_5=0, \label{ex.new}
\end{align}
with constant $d$. 
 Another unified framework including the above example has been proposed~\cite{Nishi:2016ljg}, and the Horndeski functions have been chosen as\footnote{In the previous paper~\cite{Nishi:2015pta}, $\alpha$ and $\beta$ have been introduced instead of $\beta$ and $\gamma$, respectively. However, by taking into account that $\alpha$ has already been used in Generalized Galilean Genesis, we changed the notations to avoid confusion.}
\begin{align}
G_2&=e^{2(\beta+1)\lambda\phi}g_2(Y)+e^{-2(\gamma-1)\lambda\phi}a_2(Y)\notag\\
&\quad+e^{-2(\beta+2\gamma-1)}b_2(Y),\notag\\
G_3&=e^{2\beta\lambda\phi}g_3(Y)+e^{-2\gamma\lambda\phi}a_3(Y)+e^{-2(\beta+2\gamma)\lambda\phi}b_3(Y),\notag\\
G_4&=e^{-2\gamma\lambda\phi}a_4(Y)+e^{-2(\beta+2\gamma)\lambda\phi}b_4(Y),\notag\\
G_5&=e^{-2(\beta+2\gamma+1)\lambda\phi}b_5(Y), \label{Lagr.new}
\end{align}
where 
\begin{align}
a_2(Y)&:=8\lambda^2Y(Y\partial_Y+\gamma)^2A(Y),\notag\\
a_3(Y)&:=-2\lambda(2Y\partial_Y+1)(Y\partial_Y+\gamma)A(Y),\notag\\
a_4(Y)&:=Y\partial_YA(Y),\notag\\
b_2(Y)&:=16\lambda^3Y^2(Y\partial_Y+\beta+2\gamma+1)^3B(Y),\notag\\
b_3(Y)&:=-4\lambda^2Y(2Y\partial_Y+3)(Y\partial_Y+\beta+2\gamma+1)^2B(Y),\notag\\
b_4(Y)&:=2\lambda Y(Y\partial_Y+1)(Y\partial_Y+\beta+2\gamma+1)B(Y),\notag\\
b_5(Y)&:=-(2Y\partial_Y+1)(Y\partial_Y+1)B(Y),\label{eq: a,b}
\end{align}
with arbitrary functions $A(Y)$ and $B(Y)$. Note that $g_2=-d\lambda^4-Y+\lambda^8 Y^3,\ g_3=0,\ A=-\mpl^2\lambda^8/(4Y^2),\ B=0$, and $\beta=\gamma=2$ for the above example.

Before moving to the background dynamics in the new framework, we refer to a model constructed in Ref.~\cite{Kobayashi:2016xpl}. Its Lagrangian that admits the quasi-Minkowski solution has been constructed based on the Arnowitt-Deser-Misner (ADM) formalism within the Horndeski theory. The covariantized version of the Lagrangian can be reproduced by choosing $g_2(Y), g_3(Y), A(Y)$, and $B(Y)$ in the new framework as\footnote{We changed the notation of $c$ in Ref.~\cite{Kobayashi:2016xpl} as $c\to c\lambda\sqrt{2Y_0}$. By taking $\lambda=1$ (i.e., rescaling $\phi$ to be dimensionless, $\phi\to\lambda\phi$) and replacing $\beta$ and $\gamma$ with $\alpha+\delta/2$ and $-\alpha$, respectively, one can check that the above Lagrangian is corresponding to the ADM Lagrangian in Ref.~\cite{Kobayashi:2016xpl}}
\begin{align}
g_2(Y)&=c^{-2(\beta+1)}\lambda^{-4}\biggl[-\frac{Y}{Y_0}+\frac{1}{3}\biggl(\frac{Y}{Y_0}\biggr)^2\biggr]\notag\\
&\quad\ +\frac{\sqrt{2}}{4}c^{-(2\beta+1)}\frac{(\beta-1)}{\lambda^2 Y_0^{3/2}}Y^2,\\
g_3(Y)&=\frac{3\sqrt{2}}{16}c^{-(2\beta+1)}\frac{Y}{\lambda^3 Y_0^{3/2}},\\
A(Y)&=c^{2\gamma}\lambda^{-2}\biggl[\ln\biggl(\frac{Y}{\mu^4}\biggr)-\frac{1+2\gamma}{\gamma}\biggr],\\
B(Y)&=0,
\end{align}
where $c$ is a dimensionless constant, and $\mu$ is a mass-dimension one. Note that $\gamma<0$ and $1+2\beta+4\gamma<0$ have been chosen in Ref.~\cite{Kobayashi:2016xpl}.

Under the ansatz, Eq.~(\ref{ansatz-phi}), the quasi-Minkowski solution has been obtained in the new framework as
\begin{align}
a\simeq 1+\frac{1}{2(\beta+\gamma)}\frac{\tilde h_0}{(-t)^{2(\beta+\gamma)}}=1+\mathcal{O}(Ht),
\end{align}
where
\begin{align}
H&\simeq\frac{\tilde h_0}{(-t)^{1+2(\beta+\gamma)}},\\
\beta+\gamma&>0. \label{eq: gene-cond}
\end{align}
Also, Eq.~(\ref{smallHt}) is imposed to keep the background spacetime the quasi-Minkowski one. In this framework, the Friedmann and evolution equations take the similar forms as those in Generalized Galilean Genesis:
\begin{align}
\mathcal{E}&\simeq e^{2(1+\beta)\lambda\phi}\hat\rho_2(Y_0)=0,\label{eq: new-Fried}\\
\mathcal{P}&\simeq 2\mathcal{G}_2\dot H+e^{2(1+\beta)\lambda\phi}\hat p_2(Y_0)=0, \label{eq: new-evo}
\end{align}
where 
\begin{align}
\mathcal{G}_2&:=-2e^{-2\gamma\lambda\phi}Y_0(A'+2Y_0A'')\notag\\
&\quad\ +2e^{-(\beta+2\gamma+1)\lambda\phi}H\dot\phi Y_0(6B'+9Y_0B''+2Y_0^2B'''),\\
\hat\rho_2(Y)&:=2Yg_2'-g_2-4\lambda Y(\beta g_3-Y g_3'),\label{rho2}\\
\hat p_2(Y)&:=g_2-4\beta\lambda Yg_3\notag\\
&\quad\ +8\gamma\lambda H\dot\phi e^{-2(1+\beta+\gamma)\lambda\phi}Y(A'+2YA'')\notag\\
&\quad\ -4(1+2\beta+\gamma)\lambda H^2 e^{-2(1+2\beta+2\gamma)\lambda\phi}Y^2\notag\\
&\quad\ \times(6B'+9YB''+2Y^2B'''). \label{p2}
\end{align}
The values of $Y_0$ and $\tilde h_0$ are derived from  Eqs.~(\ref{rho2}) and (\ref{p2}), respectively. The first and second terms of the evolution equation are generally quadratic in $\tilde h_0$ in the case of $B(Y)\neq0$ since $\mathcal{G}_T$ has the linear term of $h_0$ and $\hat p_2$ does the quadratic one of that. Thus, as opposed to the case of Generalized Galilean Genesis, $\tilde h_0$ is determined by solving the quadratic equation for $h_0$ in general and written as an intricate form. Whereas in the particular case where $6B'+9Y_0B''+2Y_0^2B'''=0$, the following simple expression of $\tilde h_0$ can be obtained:
\begin{align}
\tilde h_0=\frac{1}{4(1+2\beta)(2\lambda^2 Y_0)^{1+\beta+\gamma}}\frac{g_2(Y_0)-4\beta\lambda Y_0 g_3(Y_0)}{Y_0(A'+2Y_0A'')}.
\end{align}

Similarly to the case of Generalized Galilean Genesis, the background evolution is still valid even in the presence of the spatial curvature under the condition Eq.~(\ref{eq: gene-cond}) as has been shown in~\cite{Nishi:2016ljg}. Whereas one needs an additional condition in order for the spacetime anisotropies not to spoil the genesis background.
We thus briefly review the property of the spacetime anisotropy in the present framework. The ratio of the anisotropic expansion rate (denoted by $\dot\beta_{\pm}$) to the isotropic one has been obtained in the Kasner spacetime as~\cite{Nishi:2016ljg}
\begin{align}
\frac{\dot\beta_{\pm}}{H}\propto|t|^{1+2\beta}. \label{aniso1}
\end{align}
Therefore, the anisotropies do not spoil the background evolution as long as the following condition holds:
\begin{align}
\beta>-1/2. \label{aniso2}
\end{align}

 \section{Primordial Power spectra}\label{sec: power}
The scalar and tensor perturbations around the FLRW background are defined in the perturbed metric under the unitary gauge, $\delta\phi(t,\vec x)=0$, as
 \begin{align}
\D s^2=-N^2\D t^2+g_{ij}(\D x^i+N^i\D t)(\D x^j+N^j\D t), \label{eq: pert-met1}
 \end{align}
 where 
\begin{align}
N&=1+\delta n,\ N_i=\partial_i\chi,\\
g_{ij}&=a^2e^{2\zeta}\biggl(\delta_{ij}+h_{ij}+\frac{1}{2}h_{ik}h^k_j+\cdots\biggr), \label{eq: pert-met2}
\end{align}
 with $\delta n$ and $\chi$ being auxiliary fields, and we denote the curvature and tensor perturbations by $\zeta$ and $h_{ij}$, respectively. The quadratic actions for $\zeta$ and $h_{ij}$ are found, respectively, to be~\cite{Kobayashi:2011nu}
\begin{align}
S^{(2)}_\zeta&=\int{\rm d}t{\rm d}^3xa^3\biggl[\mathcal{G}_S\dot\zeta^2-\frac{\mathcal{F}_S}{a^2}(\partial_i\zeta)^2\biggr], \label{eq: quad-sc}\\
S^{(2)}_h&=\frac{1}{8}\int{\rm d}t{\rm d}^3xa^3\biggl[\mathcal{G}_T\dot h_{ij}^2-\frac{\mathcal{F}_T}{a^2}(\partial_k h_{ij})^2\biggr], \label{eq: quad-tens}
\end{align}
where the auxiliary fields were eliminated by using the constraint equations after expanding the Horndeski action up to quadratic order in the perturbations. ${\cal G}_S, {\cal F}_S, {\cal G}_T$, and ${\cal F}_T$ are defined in Appendix.~\ref{sec: App-pert}. Note that $\mathcal{G}_S, \mathcal{G}_T>0$ and $\mathcal{F}_S, \mathcal{F}_T>0$ are required to avoid the ghost and gradient instabilities, respectively.
We introduce the squared of the propagation speeds of the curvature and tensor perturbations defined, respectively, by $c_s^2:=\mathcal{F}_S/\mathcal{G}_S$ and $c_h^2:=\mathcal{F}_T/\mathcal{G}_T$.

 We also define the Fourier transform of the perturbations by
 \begin{align}
 \zeta(t,{\bf x})&=\int\frac{\D^3k}{(2\pi)^3}\tilde\zeta(t,{\bf k})e^{i{\bf k}\cdot{\bf x}},\\
 h_{ij}(t,{\bf x})&=\int\frac{\D^3k}{(2\pi)^3}\tilde h_{ij}(t,{\bf k})e^{i{\bf k}\cdot{\bf x}}.
 \end{align}
 In Fourier space, the quantized perturbations can be expanded as
 \begin{align}
\tilde\zeta(t,{\bf k})&=\zeta_{\bf k}(t)\hat a_{\bf k}+\zeta^\ast_{-{\bf k}}(t)\hat a^\dagger_{-{\bf k}},\\
\tilde h_{ij}(t,{\bf k})&=\biggl[h^{(s)}_{\bf k}\hat a^{(s)}_{\bf k}+h^{(s)\ast}_{-{\bf k}}\hat a^{(s)\dagger}_{-{\bf k}}\biggr]e^{(s)}_{ij},
 \end{align}
 where $\zeta_{\bf k}(t)$ and $h^{(s)}_{\bf k}(t)$ are the mode functions of the scalar and tensor modes, respectively, and the polarization tensor $e^{(s)}_{ij}$ satisfies
\begin{align}
    &
    \delta_{ij}e^{(s)}_{ij}({\bf k})=k^ie^{(s)}_{ij}({\bf k})=0,
    \\
    &
    e^{(s)}_{ij}({\bf k})e^{(s')\ast}_{ij}({\bf k})=\delta_{ss'}
\end{align}
with $s=\pm$ being the helicity modes of $h_{ij}$.
Here, $\hat a_{\bf k}$ ($\hat a^{(s)}_{\bf k}$) and $\hat a^\dagger_{\bf k}$ ($\hat a^{(s)\dagger}_{\bf k}$) are the creation and annihilation operators of the scalar (tensor) modes, respectively, which enjoy the canonical commutation relations:
\begin{align}
\left[\hat a_{\bf k},\hat a_{{\bf k}'}\right]&=(2\pi)^3\delta({\bf k}+{\bf k}'),\\
\biggl[\hat a^{(s)}_{\bf k},\hat a^{(s')\dagger}_{{\bf k}'}\biggr]&=(2\pi)^3\delta_{ss'}\delta({\bf k}+{\bf k}'),\\
{\rm others}&=0.
\end{align}
The equations of motion for the mode functions are derived from the quadratic actions, Eqs.~(\ref{eq: quad-sc}) and (\ref{eq: quad-tens}), as
\begin{align}
\partial_t(a^3\mathcal{G}_S\dot\zeta_{\bf k})+ak^2\mathcal{F}_S\zeta_{\bf k}&=0, \label{eq: mode-eq-s}\\
\partial_t(a^3\mathcal{G}_T\dot h^{(s)}_{\bf k})+ak^2\mathcal{F}_Th^{(s)}_{\bf k}&=0. \label{eq: mode-eq-h}
\end{align}
Now we focus on the genesis phase (i.e., $a\simeq 1$ phase) where the conformal time $\eta$ is approximately the same as the cosmic one $t$, i.e., $\eta\simeq t$.
The solutions of the mode functions are thus obtained by fixing the time dependence of $\mathcal{G}_S, \mathcal{F}_S, \mathcal{G}_T$, and $\mathcal{F}_T$. In the present paper, we fix it by requiring that the propagation speeds of the perturbations are constant for simplicity as will be argued in the following subsection.

Also, we solve Eqs.~(\ref{eq: mode-eq-s}) and~(\ref{eq: mode-eq-h}) under the initial conditions such that the solutions of the mode functions of the canonically normalized perturbations, $u_{\bf k}=\sqrt{2}a({\mathcal{G}_S\mathcal{F}_S})^{1/4}\zeta_{\bf k}$ and $v^{(s)}_{\bf k}=(a({\mathcal{G}_T\mathcal{F}_T})^{1/4}/2)h^{(s)}_{\bf k}$, in the far past coincide with those in Minkowski spacetime:
\begin{align}
\lim_{t\to-\infty}u_{\bf k}&=\frac{1}{\sqrt{2k}}e^{-ic_skt},\\
\lim_{t\to-\infty}v^{(s)}_{\bf k}&=\frac{1}{\sqrt{2k}}e^{-ic_hkt},
\end{align}
where $c_s, c_h={\rm const}$ was imposed as mentioned above. The mode functions are thus given by the positive frequency modes.

The power spectra for $\zeta$ and $h_{ij}$ are defined, respectively, by
\begin{align}
\langle\tilde\zeta({\bf k})\tilde\zeta({\bf k}')\rangle=(2\pi)^3\delta({\bf k}+{\bf k}')\frac{2\pi^2}{k^3}\mathcal{P}_\zeta,\\
\langle\tilde h_{ij}({\bf k})\tilde h_{ij}({\bf k}')\rangle=(2\pi)^3\delta({\bf k}+{\bf k}')\frac{2\pi^2}{k^3}\mathcal{P}_h,
\end{align}
where 
\begin{align}
\mathcal{P}_\zeta&:=\frac{k^3}{2\pi^2}|\zeta_{\bf k}|^2,\label{eq: power-c}\\
\mathcal{P}_h&:=\frac{k^3}{2\pi^2}\sum_{s=\pm}|h^{(s)}_{\bf k}|^2. \label{eq: power-t}
\end{align}
We also introduce the spectral indices defined by
\begin{align}
n_S-1&:=3-2|\nu_S|:=\frac{\D \ln\mathcal{P}_\zeta}{\D \ln k},\\
n_T&:=3-2|\nu_T|:=\frac{\D \ln\mathcal{P}_h}{\D \ln k}.
\end{align}
We evaluate the power spectra at the end of the genesis phase. In the usual models of inflation, the times when the phase oscillations of the mode functions stop (i.e., $-c_skt=1$ and $-c_hkt=1$) are equivalent to those when each mode crosses the Hubble (or sound) horizon (i.e., $c_sk=aH$ and $c_hk=aH$). Whereas both do not have such one-to-one relationship during the genesis phase where $a\simeq1$ and $H|t|\ll1$. We, however, call the times when $t$ enjoys $-c_skt=1$ and $-c_hkt=1$ the horizon-crossing scales for simplicity, and we evaluate the power spectra at $t=t_*$ when the perturbations are on the superhorizon scales, i.e.,  $-c_skt_*,\ -c_hkt_*\ll1$.

When $\mathcal{G}_i$ and $\mathcal{F}_i$ are of the power-law functions of $t$ as $\mathcal{G}_S,\ \mathcal{F}_S\propto|t|^{p}$ and $\mathcal{G}_T,\ \mathcal{F}_T\propto|t|^{q}$, one can derive the generic forms of the power spectra
\cite{Nishi:2016ljg}:
\begin{align}
\mathcal{P}_\zeta&=\frac{1}{8\pi^2}\frac{1}{\mathcal{F}_Sc_s}\frac{1}{t^2}\biggr|_{t=t_*}\biggl[2^{|\nu_S|-3/2}\frac{\Gamma(|\nu_S|)}{\Gamma(3/2)}\biggr]^2|c_skt_*|^{n_S},\\
\mathcal{P}_h&=\frac{2}{\pi^2}\frac{1}{\mathcal{F}_Tc_h}\frac{1}{t^2}\biggr|_{t=t_*}\biggl[2^{|\nu_T|-3/2}\frac{\Gamma(|\nu_T|)}{\Gamma(3/2)}\biggr]^2|c_hkt_*|^{n_T},
\end{align}
where
\begin{align}
\nu_S&:=\frac{1-p}{2},\\
\nu_T&:=\frac{1-q}{2}.
\end{align}

\subsection{Generalized Galilean Genesis}
In this subsection, we briefly summarize the spectral indices of the scalar and tensor power spectra in Generalized Galilean Genesis. As we explained before, it is determined by fixing the time dependence of the coefficients in the quadratic actions.

For the tensor perturbations, the time dependence has been obtained as~\cite{Nishi:2015pta}
\begin{align}
\mathcal{G}_T, \mathcal{F}_T&\simeq{\rm const.}, \label{eq: gt-ft-time-GGG}
\end{align}
giving $n_T=2$ irrespective of the concrete model.

For the scalar perturbations, the time dependence has been found to be~\cite{Nishi:2015pta}
\begin{align}
\mathcal{G}_S,\ \mathcal{F}_S\propto(-t)^{2\alpha}, \label{eq: gs-fs-time-GGG}
\end{align}
giving $\nu_S=(1-2\alpha)/2$. Thus the spectral index reads
\begin{align}
n_S-1&=\begin{cases}
\displaystyle{2\alpha+2}
 &\ \ (0<\alpha<1/2),\\
\displaystyle{4-2\alpha}
 &\ \ (\alpha>1/2).
\end{cases}
\end{align}
The two divided cases are associated with the time evolution of the superhorizon modes. In the present setup, the dominant mode of the perturbation on the superhorizon scales can be written as $\zeta_{\bf k}\propto|t|^{\nu_S-|\nu_S|}$, and thus the amplitude of the dominant mode is constant for $\nu_S>0$ while that grows for $\nu_S<0$.
The scale invariance (i.e., $n_S=1$) can be realized only for $\alpha=2$ ($\nu_S=-3/2$).

\subsection{New Framework}

First, let us focus on the tensor perturbations. The time dependence of the coefficients in the quadratic action is~\cite{Nishi:2016ljg}
\begin{align}
\mathcal{G}_T&\propto(-t)^{2\gamma}, \label{eq: gt-time-new}\\
\mathcal{F}_T&\simeq\mathcal{A}(2B'+Y_0B'')(-t)^{2(\beta+2\gamma)}+\mathcal{B}(-t)^{2\gamma}, \label{eq: ft-time-new}
\end{align}
where $\mathcal{A}$ and $\mathcal{B}$ are non-zero constants.\footnote{Strictly speaking, $\mathcal{A}$ is proportional to $1+2\beta+4\gamma$~\cite{Nishi:2016ljg} which can vanish at the present stage. However, as shown later, this cannot vanish in light of the stability conditions.} For the models with $c_h^2={\rm const}.$, 
$2B'+Y_0B''=0$ is imposed.

Then, we consider the scalar perturbations. During the genesis phase ($a\simeq 1$), $\mathcal{F}_S$ approximately takes the form
\begin{align}
\mathcal{F}_S\simeq\partial_t\biggl(\frac{\mathcal{G}_T^2}{\Theta}\biggr)-\mathcal{F}_T, \label{Fs-approx}
\end{align}
and the first term can be rewritten as
\begin{align}
\partial_t\biggl(\frac{\mathcal{G}_T^2}{\Theta}\biggr)=(1+2\beta+4\gamma)\frac{\delta}{Ht}\mathcal{G}_T\propto(-t)^{2(\beta+2\gamma)}, \label{eq: fs-time-new}
\end{align}
where 
\begin{align}
\delta:=\frac{H\mathcal{G}_T}{\Theta}={\rm const.},
\end{align}
and we used $\Theta\propto(-t)^{-(1+2\beta)}$ and $\mathcal{G}_T\propto(-t)^{2\gamma}$ which are shown in Appendix.~\ref{sec: App-pert}. The first and second terms of Eq.~(\ref{Fs-approx}) are proportional to $(-t)^{2(\beta+2\gamma)}$ and $(-t)^{2\gamma}$, respectively. Notice that the gradient instabilities occur in either the scalar or tensor perturbations if the first term vanishes, i.e., $\beta$ and $\gamma$ satisfy $1+2\beta+4\gamma=0$, since $\mathcal{F}_S=-\mathcal{F}_T$ in that case. We thus assume $1+2\beta+4\gamma\neq0$. In Ref.~\cite{Nishi:2016ljg}, by taking into account $\beta+\gamma>0$ (and also $1+2\beta+4\gamma\neq0$), the second term of Eq.~(\ref{Fs-approx}) has been ignored, which indicates that
\begin{align}
\frac{\mathcal{F}_T}{\mathcal{F}_S}=\mathcal{O}\biggl(c_h^2\frac{Ht}{\delta}\biggr)\ll1. \label{ft/fs}
\end{align}
Also, $\mathcal{G}_S$ is of the form
\begin{align}
\mathcal{G}_S=\frac{\Sigma\mathcal{G}_T^2}{\Theta^2}+3\mathcal{G}_T,\label{Gs}
\end{align}
where the first and second terms are proportional to $(-t)^{2(\beta+2\gamma)}$ and $(-t)^{2\gamma}$, respectively. As opposed to $\mathcal{F}_S$, the first term can vanish in the case of $\hat\rho'_2(Y_0)=0$ ($\Sigma\propto\hat\rho_2'(Y_0)$). The present framework has thus two divided cases about the time dependence of $\mathcal{G}_S$: $\hat\rho_2'(Y_0)=0$ and $\hat\rho_2'(Y_0)\neq0$. The latter corresponds to the class of the models with $c_s^2={\rm const}.$, and hereafter we only consider the case where
\begin{align}
\mathcal{G}_S\propto(-t)^{2(\beta+2\gamma)}. \label{eq: gs-time-new}
\end{align}

In the case of $\hat\rho_2'(Y_0)\neq0$, the second term of Eq.~(\ref{Gs}) has been ignored~\cite{Nishi:2016ljg}: $\mathcal{G}_S\simeq\Sigma\mathcal{G}_T^2/(\Theta^2)$ where the following approximation has been imposed,
\begin{align}
\frac{\mathcal{G}_T}{\mathcal{G}_S}=\mathcal{O}\biggl(c_s^2\frac{Ht}{\delta}\biggr)\ll1.\label{gt/gs}
\end{align}
We have used Eq.~(\ref{ft/fs}) to derive the above expression.

Under the conditions,  Eqs.~(\ref{ft/fs}) and (\ref{gt/gs}), one can derive the spectral indices, $n_T=3-2|\nu_T|$ and $n_S-1=3-2|\nu_S|$ with
\begin{align}
\nu_T&=\frac{1}{2}-\gamma,\\
\nu_S&=\frac{1}{2}-\beta-2\gamma.
\end{align}
The scale invariance of the scalar power spectrum can be realized for $\beta+2\gamma+1=0$ (i.e., $\nu_S=3/2$) and $\beta+2\gamma-2=0$ (i.e., $\nu_S=-3/2$).\footnote{In Refs.~\cite{Kobayashi:2016xpl,Ageeva:2018lko,Ageeva:2020buc,Ageeva:2020gti,Ageeva:2021yik}, the model with $1+2\beta+4\gamma<0$ (i.e., $\nu_S>1$) has been studied. In this parameter region, the scale-invariant scalar power spectrum is realized only from the constant mode ($\nu_S=3/2>1$).} Differently from the case of Generalized Galilean Genesis, the constant mode ($\nu_S=3/2$) yields the scale-invariant scalar power spectrum in addition to the growing one ($\nu_S=-3/2$).

\section{Strong Coupling}\label{sec: strong coupling}

Both frameworks have the parameter region where the coefficients of the quadratic actions are asymptotic to $0$ in the far past:
\begin{align}
\mathcal{G}_S, \mathcal{F}_S&\propto|t|^{p},\label{st-GSFS}\\ \mathcal{G}_T, \mathcal{F}_T&\propto|t|^{q}, \label{st-GTFT}
\end{align}
with $p$ and/or $q$ being negative. A concrete model having those asymptotic properties was constructed in Ref.~\cite{Kobayashi:2016xpl}. 
Eqs.~(\ref{st-GSFS}) and (\ref{st-GTFT}) imply that couplings of non-linear interactions for the canonically normalized perturbations naively diverge, and thus strong coupling occurs. Also, even if $\mathcal{G}_S,\ \mathcal{F}_S,\ \mathcal{G}_T$, and $\mathcal{F}_T$ increase as the time goes back (i.e., $p, q>0$), strong coupling can occur if couplings of higher-order interactions, e.g., the cubic interactions, diverge. In the present paper, if the canonically normalized perturbations are strongly coupled at cubic order in the far past, we require, similarly to Refs.~\cite{Ageeva:2018lko,Ageeva:2020buc,Ageeva:2020gti,Ageeva:2021yik,Ageeva:2022fyq,Ageeva:2022asq}, that the classical energy scale of the genesis background 
$E_*$ (Max\{$H,\dot H^{1/2}, \dot H/H$, {\rm etc}\}) is much lower than the scale $\Lambda$ at which strong coupling occurs:
\begin{align}
E_*\sim\frac{1}{t}\ll\Lambda,\label{eq: cond-st}
\end{align}
where we used $H\ll\dot H^{1/2}\ll\dot H/H\sim1/t$. We stress that the above scaling also corresponds to the frequencies of the perturbations at the horizon-crossing scale. Therefore, Eq.~(\ref{eq: cond-st}) would be also necessary to avoid the strong coupling problem around the horizon-crossing scale.

\subsection{Generalized Galilean Genesis}

We first write down the cubic interaction terms of the curvature perturbation. The components of the cubic interactions are obtained as
\begin{align}
\mathcal{L}^{(3)}_\zeta&\supset(-t)^{1+6\alpha}(\partial_t)^3\zeta^3,\ (-t)^{4\alpha}(\partial_t)^2\zeta^3,\notag\\
&\quad\  (-t)^{2(1+3\alpha)}(\partial_t)^2(\partial_i)^2\zeta^3, (-t)^{1+4\alpha}(\partial_t)(\partial_i)^2\zeta^3,\notag\\ 
&\quad\ (-t)^{3(1+2\alpha)}(\partial_t)(\partial_i)^4\zeta^3,\ (-t)^{0}(\partial_i)^2\zeta^3,\notag\\ 
&\quad\ (-t)^{2(1+2\alpha)}(\partial_i)^4\zeta^3, \label{eq: cubic-s-components-GGG}
\end{align}
where we used the cubic action summarized in Appendix.~\ref{sec: App-pert} and also Eqs.~(\ref{eq: gt-ft-time-GGG}),~(\ref{eq: gs-fs-time-GGG}), (\ref{eq: sigma-ggg}), (\ref{eq: theta-ggg}), (\ref{eq: Xi-ggg}), and~(\ref{eq: Gamma-ggg}). After a change of a variable from the original variable, $\zeta$, to the canonically normalized one, $u=\sqrt{2}a(\mathcal{G}_S\mathcal{F}_S)^{1/4}\zeta$, we obtain the conventional form of the cubic action for the canonically normalized curvature perturbation as
\begin{align}
\mathcal{L}^{(3)}_u&=\frac{1}{{\Lambda^s
_1}^2}(\partial_t)^3u^3+\frac{1}{\Lambda^s_2}(\partial_t)^2u^3\notag\\
&\quad +\frac{1}{{\Lambda^s_3}^3}(\partial_t)^2(\partial_i)^2u^3+\frac{1}{{\Lambda^s_4}^2}(\partial_t)(\partial_i)^2u^3\notag\\
&\quad +\frac{1}{{\Lambda^s_5}^4}(\partial_t)(\partial_i)^4u^3+\frac{1}{\Lambda^s_6}(\partial_i)^2u^3+\frac{1}{{\Lambda^s_7}^3}(\partial_i)^4u^3, \label{eq: conv-sss}
\end{align}
where $\Lambda^s_i~(i=1,\cdots,7)$ characterize the strong coupling scales of the interaction terms. Each scale evolves in time as
\begin{align}
\Lambda^s_1&\propto(-t)^{-(1+3\alpha)/2},\ \Lambda^s_2\propto(-t)^{-\alpha},\notag\\
\Lambda^s_3&\propto(-t)^{-(2+3\alpha)/3},\ \Lambda^s_4\propto(-t)^{-(1+\alpha)/2},\notag\\
\Lambda^s_5&\propto(-t)^{-(3+3\alpha)/4},\ \Lambda^s_6\propto(-t)^{3\alpha},\notag\\ \Lambda^s_7&\propto(-t)^{-(2+\alpha)/3}. \label{eq: cano-cubic-s-components-GGG}
\end{align}
Here, $\Lambda^s_i$ ($i\neq6$) are always asymptotic to $0$ in the far past since $\alpha>0$, and hence we impose Eq.~(\ref{eq: cond-st}) on $\Lambda^s_i$ ($i\neq6$) to avoid strong coupling. By parametrizing the time dependence of $\Lambda^s_i$ as $\Lambda^s_i\propto(-t)^{-x_i}$, we obtain the following conditions:
\begin{align}
1>x_i.
\end{align}
In particular, the condition obtained from the $\Lambda^s_1$-term (i.e., $1>x_1$) reads
\begin{align}
\alpha<1,
\end{align}
which is incompatible with the condition for the scalar power spectrum to be nearly scale-invariant, i.e., $\alpha\simeq2$. In this framework, the strong coupling problem is thus unavoidable if one requires the scale-invariant power spectrum.

\subsection{New Framework}

By following the same procedure as in the previous subsection, we can obtain the conventional form of the cubic action of the canonically normalized perturbation. In general, the most dangerous terms in the cubic action are
\begin{align}
\mathcal{L}_{sss}&\supset \frac{2\mu\mathcal{G}_S^2}{\mathcal{G}_T^2}\dot\zeta^3,\frac{\Gamma\mathcal{G}_S^2}{2\Theta\mathcal{G}_T}\dot\zeta^3 ,\\ \label{eq: danger}
&\Rightarrow \frac{1}{\Lambda^2}(\partial_t)^3u^3\ {\rm with}\ \Lambda\propto(-t)^{-(1+3\beta+2\gamma)/2},
\end{align}
where we used Eqs.~(\ref{eq: gt-time-new}), (\ref{eq: gs-time-new}), (\ref{eq: Gamma-new}), (\ref{eq: theta-new}), and~(\ref{eq: def-mu}) to derive the time dependence of $\Lambda$.
By requiring Eq.~(\ref{eq: cond-st}), one obtains
\begin{align}
    1-3\beta-2\gamma>0,
\end{align}
 which indicates $|\nu_S|<3/2$. Thus the spectral index of the scalar power spectrum is blue. This is conflicted with the Planck results, $n_S\simeq0.96$ (i.e., $|\nu_S|>3/2$). Note that by replacing $\beta$ and $\gamma$ with $\beta\to\alpha+\delta/2$ and $\gamma\to-\alpha$, respectively, and taking $B(Y)=0$ (and hence $\mu=0$), one can check that our result in the case of $\mu=0$ and $\Gamma\neq0$, $1-3\beta-2\gamma>0$, can reproduce $2-3\delta-2\alpha>0$ which has been obtained as the strongest constraint on the parameters in the previous papers~\cite{Ageeva:2018lko,Ageeva:2020buc,Ageeva:2020gti,Ageeva:2021yik}.

Then, we impose both $\mu=0$ and $\Gamma=0$. One can find from Eqs.~(\ref{Lagr.new}), (\ref{eq: a,b}), and~(\ref{eq: def-mu}) that $\mu$ vanishes for $6B'+9Y_0 B''+2 Y_0^2 B'''=0$. The explicit form of $\Gamma$ is generally intricate due to the presence of $H$, and $\Gamma=0$ is realized by choosing $A(Y),~B(Y),~ g_2(Y)$, and $g_3(Y)$ appropriately. We here emphasize that the Lagrangian in Refs.~\cite{Kobayashi:2016xpl, Ageeva:2018lko, Ageeva:2020buc, Ageeva:2020gti, Ageeva:2021yik} where $G_4=G_4(\phi)$ and $G_5=0$ does not admit $\Gamma=0$ since $\Gamma=\mathcal{G}_T>0$ in that model. In the present case, the components of the cubic action read
\begin{align}
\mathcal{L}_{sss}&\supset (-t)^{1+4\beta+6\gamma}(\partial_t)^3\zeta^3,\ (-t)^{2(\beta+2\gamma)}(\partial_t)^2\zeta^3,\notag\\
&\quad\ (-t)^{2(1+2\beta+3\gamma)}(\partial_t)^2(\partial_i)^2\zeta^3,\notag\\
&\quad\ (-t)^{1+4\beta+6\gamma}(\partial_t)(\partial_i)^2\zeta^3,\ (-t)^{2\gamma}(\partial_i)^2\zeta^3,\notag\\
&\quad\ (-t)^{2(1+2\beta+3\gamma)}(\partial_i)^4\zeta^3, \label{eq: cubic-s-components-new}
\end{align}
where we used the cubic action in Appendix.~\ref{sec: App-pert} and also Eqs.~(\ref{eq: gt-time-new}), (\ref{eq: ft-time-new}), (\ref{eq: gs-time-new}), (\ref{eq: fs-time-new}), (\ref{eq: sigma-new}), (\ref{eq: theta-new}), (\ref{eq: Xi-new-time}), and~(\ref{eq: Gamma-new}).
Imposing Eq.~(\ref{eq: cond-st}) on the above interaction terms yields
\begin{align}
1-\beta>0. \label{nost-sss-new}
\end{align} 
In contrast to the general case (i.e., $\mu\neq0$ or $\Gamma\neq0$), the slightly red scalar power spectrum is still allowed.

\begin{figure}[htb]
\begin{center}
\includegraphics[width=70mm]{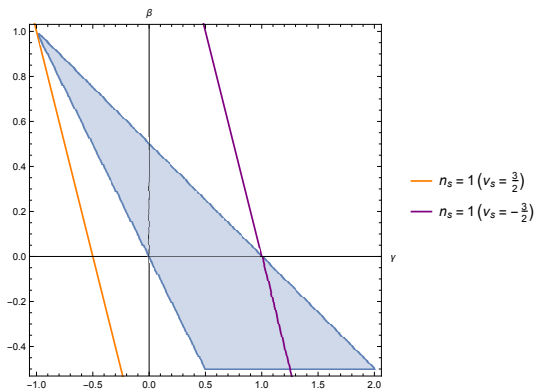}
\caption{The allowed parameter region in the $\beta-\gamma$ plane where $\mu=0=\Gamma$. The orange and purple lines denote the parameter region in which the scale invariant curvature perturbation is generated from the constant and growing modes, respectively. The blue shaded region is the parameter region in which the strong coupling can be avoided.} \label{fig1}
\end{center}
\end{figure}

Strong coupling can also occur in the cross-interactions among the scalar and tensor perturbations and the self-interaction among the tensor perturbations. Therefore, we analyze all of the other cubic interactions of the scalar and tensor perturbations. The arguments are parallel to the previous ones, and thus we show only the results which are obtained by using Eqs.~(\ref{eq: cond-st}) and~(\ref{eq: strong-new-cross}). The resultant model space avoiding strong coupling is plotted in Fig.~\ref{fig1}. In the parameter space, the range of $\beta$ is $-1/2<\beta<0$. (The lower bound is determined from the argument on the spacetime anisotropy, Eq.~(\ref{aniso2}), and the upper one is from that on strong coupling.) Note that the model parameters which can yield the scale invariance of the scalar power spectrum are located at the edge of the parameter region without strong coupling. The parameter region realizing $n_S\simeq0.96$ from the constant mode (i.e., $\nu_S\simeq3/2$) and avoiding strong coupling are not overlapped. 
In the viable model space, only the models having the growing mode (i.e., $\nu_S\simeq-3/2$) can enjoy $n_S\simeq0.96$, and the primordial power spectra of the curvature and tensor perturbations can be obtained, respectively, as
\begin{align}
\mathcal{P}_\zeta&\simeq\frac{1}{8\pi^2}\frac{1}{\mathcal{F}_Sc_s}\frac{1}{t^2}\biggr|_{t=t_*},\\
\mathcal{P}_h&=\frac{2}{\pi^2}\frac{1}{\mathcal{F}_Tc_h}\frac{1}{t^2}\biggr|_{t=t_*}\biggl[2^{|\nu_T|-3/2}\frac{\Gamma(|\nu_T|)}{\Gamma(3/2)}\biggr]^2|c_hkt_*|^{n_T},
\end{align}
where $\nu_T=(\beta-1)/2$ and $n_T=2+\beta$ with $-1/2<\beta<0$. Then, the tensor-to-scalar ratio reads
\begin{align}
r:=\frac{\mathcal{P}_h}{\mathcal{P}_\zeta}\simeq16\frac{\mathcal{F}_S}{\mathcal{F}_T}\frac{c_s}{c_h}\biggr|_{t=t_*}\biggl[2^{|\nu_T|-3/2}\frac{\Gamma(|\nu_T|)}{\Gamma(3/2)}\biggr]^2|c_hkt_*|^{n_T}.
\end{align}
Now, by recalling Eq.~(\ref{ft/fs}), one can find that the tensor-to-scalar ratio is enhanced by $\mathcal{F}_S/\mathcal{F}_T$. However, the tensor tilt is always blue: $3/2<n_T<2$, potentially leading to a small tensor-to-scalar ratio due to the suppression factor, $|c_hkt_*|^{n_T}\ll1$. More explicitly, one has
\begin{align}
r\propto\frac{\mathcal{F}_S}{\mathcal{F}_T}\biggl(10^{-59}Ht_*\times\frac{\mpl}{H}\biggr)^{n_T}\biggr|_{t=t_*}\biggl(\frac{k}{0.002 {\rm Mpc}^{-1}}\biggr)^{n_T}.
\end{align}
By inspecting the above at $k=0.002 {\rm Mpc}^{-1}$ to compare with the Planck results~\cite{Planck:2018jri}, one can find that the tensor-to-scalar ratio can be much smaller than unity for the models in which $H/\mpl$ is sufficiently larger than $10^{-59}H|t|(\ll10^{-59})$ at the end of the genesis phase. 
Note that the blue-tilted tensor power spectrum with $n_T=\mathcal{O}(1)$ makes the detection of the primordial gravitational waves on the CMB scales challenging. In the following section, we investigate the scalar non-Gaussianity to discuss the observational consistency.

The above argument on the strong coupling problem might be insufficient to conclude the presence or absence of strong coupling. First, there is a possibility that strong coupling occurs in higher-order interactions. It is hence not evident that the analysis of strong coupling at higher-order interactions does not yield tighter constraints on the model parameters. Nevertheless, in Ref.~\cite{Ageeva:2020buc}, it has been shown in the subclass of the new framework that the strongest condition is obtained from the argument at cubic order. Therefore, it would be important to ascertain the extent to which this statement generally holds. Second, to verify that the strong coupling problem is indeed absent at the onset of genesis, one needs to consider the scattering process inside the horizon (see further discussions~\cite{ Ageeva:2022fyq, Ageeva:2022asq,Cai:2022ori}). 
The consideration of the scattering process would clarify how reasonable the present naive argument is. Furthermore, in order to justify the use of the perturbation theory completely, one would also need to take account of loop corrections (see e.g., Refs.~\cite{Leblond:2008gg,Senatore:2009cf}).
We leave these points to the future work~\cite{Akama:prep}, and we have roughly evaluated Eq.~(\ref{strong-estimate}) in the present paper.

Before closing this section, we comment on a no-go theorem for non-singular cosmologies found in the previous papers~\cite{Libanov:2016kfc,Kobayashi:2016xpl,Cai:2016thi,Creminelli:2016zwa} in which it has been shown that non-singular cosmological solutions (i.e., $a(t)>0$ during the entire history) in the Horndeski theory are plagued with some no-go theorem. This theorem states that the curvature perturbations suffer from the gradient instabilities (i.e., $\mathcal{F}_S<0$) unless the following integral converges in the past infinity ($t_i=-\infty$) and/or future one ($t_f=+\infty$):
\begin{align}
\int^{t_f}_{t_i} a(t')\mathcal{F}_T(t')\D t'. \label{no-go-int}
\end{align}
As the non-singular cosmological model to avoid the gradient instabilities, the case which $\mathcal{F}_T$ converges in the past infinity has been considered so far~\cite{Kobayashi:2016xpl, Ageeva:2018lko, Ageeva:2020buc, Ageeva:2020gti, Ageeva:2021yik, Ageeva:2022fyq, Ageeva:2022asq}. 
 By recalling $\mathcal{F}_T\propto(-t)^{2\gamma}$, the models satisfying 
\begin{align}
1+2\gamma<0 \label{eq: nogo-1}
\end{align}
correspond to the case. For the models with $\beta+2\gamma=2$ (i.e., the scale-invariant power spectrum of the curvature perturbation), Eq.~(\ref{eq: nogo-1}) is equivalent to
\begin{align}
\beta>3. \label{eq: nogo-2}
\end{align}
Eq.~(\ref{eq: nogo-2}) is incompatible with the result of the analysis of the strong coupling since $-1/2<\beta<0$ in the allowed region. Therefore, the examples avoiding the gradient instabilities cannot realize the avoidance of strong coupling and the scale invariance of the scalar power spectrum simultaneously. One of the ways to overcome this is to invoke beyond Horndeski terms at somewhere during the entire history.\footnote{It has been found in the context of the second-order scalar-tensor theories that some higher-derivative extensions of the Horndeski theory or an introduction of a cuscuton field~\cite{Afshordi:2006ad,Afshordi:2007yx} are required to avoid the no-go theorem~\cite{Cai:2016thi,Creminelli:2016zwa,Cai:2017tku,Cai:2017dyi,Kolevatov:2017voe,Boruah:2018pvq,Ye:2019frg,Ye:2019sth,Quintin:2019orx}. (See Refs.~\cite{Creminelli:2016zwa} and~\cite{Akama:2017jsa} for the no-go arguments in the presence of multiple scalar fields, and also~\cite{Akama:2018cqv} and~\cite{Heisenberg:2018wye} for those in the presence of the spatial curvature and vector modes, respectively.)} In the present paper, we evade the no-go theorem by supposing that the quasi-Minkowski phase which we are focusing on is described by the Horndeski theory and some beyond Horndeski terms are developed at some regime away from the genesis phase in the entire cosmic expansion history. Under this assumption, we do not invoke any beyond Horndeski terms in the present paper and we still continue to analyze the primordial non-Gaussianities generated during the genesis phase in the present setup.

\section{Primordial non-Gaussianity}\label{sec: non-G}

In this section, we compute the primordial non-Gaussianity generated from the viable models which realize the slightly red scalar power spectrum and have no strong coupling. In doing so, we use the mode function in the case of the scale-invariant power spectrum generated from the growing mode and eliminate $\gamma$ by using the condition for the scale invariance, $\beta+2\gamma-2=0$.
Following the in-in formalism, the three-point correlation function at the end of the genesis phase can be computed as
\begin{align}
&\langle\hat\zeta({\bf k}_1)
\hat \zeta({\bf k}_2)\hat \zeta({\bf k}_3)\rangle\notag\\
&=-i\int^{t_\ast}_{-\infty}\D t'a(t)\langle[\hat\zeta(t_\ast,{\bf k}_1)
\hat\zeta(t_\ast,{\bf k}_2)\hat\zeta(t_\ast,{\bf k}_3),H_{{\rm int}}(t')]\rangle, \label{InIn}\\
&=:(2\pi)^7\delta\left({\bf k}_1+{\bf k}_2+{\bf k}_3\right)\frac{\mathcal{P}_\zeta^2}{k_1^3k_2^2k_3^3}\mathcal{A_\zeta},
\end{align}
where $H_{\rm int}$ is the interaction Hamiltonian of the curvature perturbation defined by
\begin{align}
H_{\rm int}=-\int \D^3x \mathcal{L}^{(3)}_{\zeta}.
\end{align}
Here, ${\cal L}_\zeta^{(3)}$ denotes
the cubic Lagrangian of the curvature perturbation which has been obtained as~\cite{Gao:2011qe,DeFelice:2011uc,Gao:2012ib}
\begin{align}
{\cal L}_\zeta^{(3)}&=a^3{\mathcal G}_S
\biggl\{\frac{\Lambda_1}{H}
\dot\zeta^3+\Lambda_2\zeta\dot\zeta^2+\Lambda_3\zeta\frac{\left(\partial_i\zeta\right)^2}{a^2}
\notag\\
&\quad\quad\quad +\frac{\Lambda_4}{H^2}
\dot\zeta^2\frac{\partial^2\zeta}{a^2}+\Lambda_5\dot\zeta\partial_i\zeta\partial_i
\psi+\Lambda_6\partial^2\zeta\left(\partial_i\psi\right)^2\notag\\
&\quad\quad\quad
+\frac{\Lambda_7}{H^2}
\frac{1}{a^4}\left[\partial^2\zeta\left(\partial_i\zeta\right)^2-\zeta\partial_i\partial_j\left(\partial_i\zeta\partial_j\zeta\right)\right]\notag\\
&\quad\quad\quad+\frac{\Lambda_8}{H}\frac{1}{a^2}\left[\partial^2\zeta\partial_i\zeta\partial_i\psi-\zeta\partial_i\partial_j\left(\partial_i\zeta\partial_j\psi\right)\right]\biggr\}\notag\\
&\quad+F(\zeta)E_S, \label{eq: cubic}
\end{align}
where
\begin{align}
F(\zeta)&:=-\frac{\Lambda_{{\rm red},1}}{H}\zeta\dot\zeta-\frac{\Lambda_{{\rm red},2}}{H}(\partial_i\zeta\partial_i\psi-\partial^{-2}\partial_i\partial_j(\partial_i\zeta\partial_j\psi))\notag\\
&\quad\ +\frac{\Lambda_{{\rm red},3}}{a^2H^2}((\partial_i\zeta)^2-\partial^{-2}\partial_i\partial_j(\partial_i\zeta\partial_j\zeta)),\label{Fzeta}\\
E_S&:=-2\partial_t(a^3\mathcal{G}_S\dot\zeta)+2a\mathcal{F}_S\partial^2\zeta.
\end{align}
Each $\Lambda_i$ is defined in Appendix.~\ref{sec: App-pert}. The explicit forms of $\Lambda_{{\rm red},i}$ are given by
\begin{align}
\Lambda_{{\rm red},1}&=\frac{H\mathcal{G}_T\mathcal{G}_S}{\Theta\mathcal{F}_S},\\
\Lambda_{{\rm red},2}&=\frac{H\Gamma\mathcal{G}_S}{2\Theta\mathcal{G}_T},\\
\Lambda_{{\rm red},3}&=\frac{H^2\Gamma\mathcal{G}_T}{4\Theta^2},
\end{align}
and hence $\Lambda_{{\rm red},(2,3)}=0$ in the allowed parameter region. Note that $E_S=0$ is the equation of motion for the linear curvature perturbation. One can eliminate the last line of Eq.~(\ref{eq: cubic}) by performing the following field redefinition:
\begin{align}
\zeta\to\zeta-F(\zeta).
\end{align}
In light of that $\dot\zeta\simeq-3\zeta/t$ on the superhorizon scales, the field redefinition in Fourier space reduces to
\begin{align}
\zeta({\bf k})\to\zeta({\bf k})-\frac{3\Lambda_{{\rm red},1}}{Ht}\int\frac{{\rm d}^3k'}{(2\pi)^3}\zeta({\bf k}')\zeta({\bf k}-{\bf k}').\label{field}
\end{align}
The resultant form of $\mathcal{A}_\zeta$ reads
\begin{widetext}
\begin{align}
\mathcal{A}_\zeta&=
    \frac{3}{2}\biggl[\frac{3}{4}\frac{\Lambda_{1,0}}{Ht}-\frac{9}{2(\beta-4)}\frac{\Lambda_{1,1}}{Ht}\biggr]\biggr|_{t=t_*}\sum_i k_i^3+\biggl[-\frac{1}{8}\Lambda_{2,0}+\frac{3(1+2\beta)}{4(\beta-1)(\beta-4)}\Lambda_{2,1}\biggr]\biggl. \biggr|_{t=t_*}\sum_i k_i^3
\notag\\
&\quad\ +\frac{3}{16(\beta-4)(\beta-1)}\Lambda_{5,0}|_{t=t_*}\frac{1}{k_1^2k_2^2k_3^2}\biggl[(\beta-4)\biggl(\sum_{i\neq j}k_i^7k_j^2-\sum_{i\neq j}k_i^5k_j^4\biggr)-2(1+2\beta)k_1^2k_2^2k_3^2\sum_i k_i^3\biggr]
\notag\\
    &\quad +\frac{3}{8(\beta-4)(\beta-1)}\Lambda_{6,0}|_{t=t_*}\frac{1}{k_1^2k_2^2k_3^2}\biggl[3\sum_ik_i^9-3\sum_{i\neq j}k_i^7 k_j^2+(\beta-1)\sum_{i\neq j}k_i^6k_j^3-(\beta-1)\sum_{i\neq j}k_i^5k_j^4
\notag\\
    &\quad -(\beta-1)k_1^2k_2^2k_3^2\sum_{i\neq j}k_i^2 k_j\biggr]-\frac{3}{2}\frac{\Lambda_{{\rm red},1}}{Ht}\biggl|_{t=t_*}\sum_i k_i^3, \label{non-G-amp-G0}
\end{align}
\end{widetext}
where we ignored the terms proportional to the positive powers of $|c_sKt_*|$ which are suppressed on the superhorizon scales, $|c_sKt_*|\ll1$. One here notices that the amplitude of the non-Gaussianity is generically enhanced in proportion to $|Ht|^{-1}\gg1$. This factor is of $\mathcal{O}(\epsilon)$ where $\epsilon$ is the usual “slow-roll" parameter defined by $\epsilon:=-\dot H/H^2$. In the case of inflation and matter bounce cosmology which generate the scale-invariant scalar power spectrum, one has $\epsilon\ll1$ and $\epsilon=\mathcal{O}(1)$, respectively. Whereas in the present case, $\epsilon$ is much larger than unity. This fact can result in an enhanced non-Gaussianity, compared to the non-Gaussianities from those models. Below we explore the possibility to suppress this enhancement by detuning model parameters. In doing so, we compare the non-Gaussian amplitude with the observational constraint on that obtained by Planck~\cite{Planck:2019kim}, and we evaluate the following non-linearity parameter
\begin{align}
    f_{\rm NL} := \frac{10}{3}\frac{\cal A_\zeta}{\sum_i k_i^3},
\end{align}
at the squeezed, equilateral, and folded limits which are defined by $k_1\ll k_2=k_3$, $k_1=k_2=k_3$, and $k_1=2k_2=2k_3$, respectively.
The non-linear parameters at these limits are denoted by $f_{\rm NL}^{\rm local}$, $f_{\rm NL}^{\rm equil}$, and $f_{\rm NL}^{\rm fold}$, respectively.

Each of the non-linearity parameters is obtained from Eq.~(\ref{non-G-amp-G0}) as
\begin{align}
f^{\rm local}_{\rm NL}&=\mathcal{C}+\frac{5(2-5\beta)}{16(\beta-4)(\beta-1)}\frac{k_1^2}{k_2^2}\frac{\mathcal{G}_S}{\mathcal{G}_T}\biggr|_{t=t_*},\\
f^{\rm equil}_{\rm NL}&=\mathcal{C}-\frac{5(1+2\beta)}{16(\beta-4)(\beta-1)}\frac{\mathcal{G}_S}{\mathcal{G}_T}\biggr|_{t=t_*},\\
f^{\rm fold}_{\rm NL}&=\mathcal{C}+\frac{94-\beta}{4(\beta-4)(\beta-1)}\frac{\mathcal{G}_S}{\mathcal{G}_T}\biggr|_{t=t_*},
\end{align}
where $\mathcal{C}$ denotes the terms contributed from the terms $\propto\sum_i k_i^3$ in Eq.~(\ref{non-G-amp-G0}). To simplify the argument on the consistency with the current constraints on $f_{\rm NL}$, we consider the difference of each $f_{\rm NL}$ instead of the magnitude of itself. In doing this, the above common contributions do not affect that argument, and hence we do not write the explicit form of $\mathcal{C}$. 
The difference of each non-linearity parameter reads
\begin{align}
f^{\rm local}_{\rm NL}-f^{\rm fold}_{\rm NL}&\simeq-\frac{94-\beta}{4(\beta-4)(\beta-1)}\frac{\mathcal{G}_S}{\mathcal{G}_T}\biggr|_{t=t_*},\\
f^{\rm local}_{\rm NL}-f^{\rm equil}_{\rm NL}&\simeq\frac{5(1+2\beta)}{16(\beta-4)(\beta-1)}\frac{\mathcal{G}_S}{\mathcal{G}_T}\biggr|_{t=t_*},\\
f^{\rm equil}_{\rm NL}-f^{\rm fold}_{\rm NL}&=-\frac{3(2\beta+127)}{16(\beta-4)(\beta-1)}\frac{\mathcal{G}_S}{\mathcal{G}_T}\biggr|_{t=t_*},
\end{align}
where we used $k_1/k_2\ll1$. As long as $\beta$ enjoys $-1/2<\beta<0$, the above three quantities are generally much larger than unity. Therefore, at least two out of the non-linearity parameters are much larger than unity, which indicates that the genesis models are ruled out by the constraints on the primordial scalar non-Gaussianity~\cite{Planck:2019kim}. One may consider the $1+2\beta\to0$ limit, but at least one of the non-linearity parameters is much larger than unity in such a case, and thus that case is also unacceptable from the observational viewpoint. 

The above arguments imply that the dangerous terms in $f_{\rm NL}$ are naively at most of $\mathcal{O}(\mathcal{G}_S/\mathcal{G}_T)|_{t=t_*}\sim\mathcal{O}(|Ht|^{-1})|_{t=t_*}$. To check the validity of the perturbation theory from both theoretical and observational viewpoints, we estimate $f_{\rm NL}\zeta$ which is roughly the ratio of the cubic action to the quadratic one (see e.g., Refs.~\cite{Leblond:2008gg,Baumann:2011dt,Joyce:2011kh}). This quantity contributed from the dangerous terms can be evaluated as
\begin{align}
f_{\rm NL}\zeta\sim \frac{\zeta}{Ht}\Bigg|_{t=t_*}. \label{strong-estimate}
\end{align}
The amplitude of the curvature perturbation has been constrained as $\zeta(t_*)\sim\mathcal{O}(10^{-5})$. For the models enjoying $f_{\rm NL}\zeta|_{t=t_*}\ll1$, say $10^{-3}\lesssim H|t|\ll1$ at $t=t_*$, the perturbative analysis is still valid at the end of the genesis phase, even though those are ruled out by the arguments on $f_{\rm NL}$. Also, after the horizon-crossing, Eq.~(\ref{strong-estimate}) is proportional to $(Ht^4)^{-1}\propto (-t)^{\beta-1}$ which monotonically increases with time in the allowed parameter region (i.e., for $-1/2<\beta<0$). Therefore, at least based on the present naive arguments, the perturbative analysis would be guaranteed from the time of the horizon-crossing to that of the end of the genesis phase as long as it is guaranteed at the latter time.

\section{Summary}\label{sec: summary}
In the first half of the present paper, we have derived the conditions to avoid the strong coupling problem in two unified frameworks of Galilean Genesis in which the propagation speeds of the perturbations are constant. As a result, we have clarified that the new framework has the parameter region without strong coupling and with the slightly red-tilted scalar power spectrum. This is in contrast to the case of the model studied in Refs.~\cite{Ageeva:2018lko,Ageeva:2020gti,Ageeva:2020buc,Ageeva:2021yik} where the spectral index of the scalar power spectrum was blue in the model space without strong coupling.  In general, the Lagrangian of the new framework includes the functional degrees of freedom ($g_2(Y), g_3(Y), A(Y)$, and $B(Y)$, etc) to reduce the dangerous cubic interaction terms leading to strong coupling. We particularly chose the general class of the models which allow $\mu$ and $\Gamma$ to vanish thanks to the functional degrees of freedom. (Note that $\mu\neq0$ and/or $\Gamma\neq0$ yield the most dangerous cubic interaction terms.) Whereas the theory in Refs.~\cite{Kobayashi:2016xpl,Ageeva:2018lko,Ageeva:2020gti,Ageeva:2020buc,Ageeva:2021yik} where $\mu=0$ does not allow $\Gamma=0$. This difference has yielded the parameter region without strong coupling and with the slightly red power spectrum.

In the second half of the present paper, we have calculated the non-Gaussianity of the curvature perturbations. The primordial bispectrum has been calculated for the first time in the context of Galilean Genesis in the present paper. Note that the trispectrum has been studied in Ref.~\cite{Libanov:2011bk}. 
 We have also found that the non-Gaussian amplitudes are generically enhanced and clarified that the non-Gaussian signatures are different from those in the other scenarios generating the scale-invariant power spectrum because of the large value of the usual slow-roll parameter, $\epsilon=-\dot H/H^2\gg1$. Then, by evaluating the non-linearity parameters, we have shown that the models considered in the present paper cannot be consistent with the current CMB data due to the enhanced non-Gaussian amplitudes. 

As the further study of various aspects of the perturbative analysis in the context of Galilean Genesis, it would be important to investigate the scattering process and loop corrections as has been mentioned in Sec.~\ref{sec: strong coupling}. 
It would be also interesting to go outside the model space considered in the present paper, e.g., by allowing the time variations of the propagation speeds of the perturbations and using beyond Horndeski theories~\cite{Kobayashi:2015gga,Cai:2017tku,Mironov:2019qjt, Mironov:2019haz,Volkova:2019jlj,Ilyas:2020zcb,Zhu:2021ggm,Cai:2022ori}.

\acknowledgments
We would like to thank Tsutomu Kobayashi for collaboration in the early stage of this work and for his helpful comments on the manuscript. We also thank Sakine Nishi for collaboration in the early stage. We thank Tomohiro Fujita, Yuichiro Tada, Daisuke Yamauchi, and Shuichiro Yokoyama for fruitful discussions. 
The work of SA was supported by the grant No. UMO-2018/30/Q/ST9/00795 from the National Science Center, Poland and was partially supported by the JSPS Research Fellowships for Young Scientists
No.~18J22305 in the early stage of his work. The work of SH was supported by the JSPS KAKENHI Grant No. JP21H01080.


\appendix
\begin{widetext}

\section{Background Equations}\label{sec: background}
From the Horndeski action, Eq.~(\ref{eq: Horndeski action}), one can obtain the Friedmann and evolution equations, respectively, as
\begin{align}
\mathcal{E}&=2XG_{2X}-G_2+6X\dot\phi{H}G_{3X}-2XG_{3\phi}-6H^2G_4+24H^2{X}\left(G_{4X}+XG_{4XX}\right)\notag\\
&\quad-12HX\dot\phi{G_{4\phi{X}}}-6H\dot\phi{G_{4\phi}}+2H^3X\dot\phi\left(5G_{5X}+2XG_{5XX}\right)-6H^2X\left(3G_{5\phi}+2XG_{5\phi{X}}\right),\\
\mathcal{P}&=G_2-2X(G_{3\phi}+\ddot{\phi}G_{3X})+2(3H^2+2\dot{H})G_4-12H^2XG_{4X}-4H\dot{X}G_{4X}-8\dot{H}XG_{4X}\notag\\
&\quad-8HX\dot{X}G_{4XX}+2(\ddot{\phi}+2H\dot{\phi})G_{4\phi}+4XG_{4\phi\phi}+4X(\ddot{\phi}-2H\dot{\phi})G_{4\phi{X}}-2X(2H^3\dot{\phi}+2H\dot{H}\dot{\phi}+3H^2\ddot{\phi})G_{5X}
\notag\\
&\quad-4H^2X^2\ddot{\phi}G_{5XX}+4HX(\dot{X}-HX)G_{5\phi{X}}+2\left[2(HX)^{{\boldsymbol \cdot}}+3H^2X\right]G_{5\phi}+4HX\dot{\phi}G_{5\phi\phi}.
\end{align}
In Generalized Galilean Genesis, the Friedmann and evolution equations have been obtained as~\cite{Nishi:2015pta}
\begin{align}
\mathcal{E}&=e^{2(\alpha+1)\lambda\phi}\hat\rho_1(Y)+6H\dot\phi e^{2\alpha\lambda\phi}\left[Yg_3'-2\alpha\lambda g_4+2(3-2\alpha)\lambda Y g_4'+4\lambda Y^2 g_4''\right]\notag\\
&\quad -3H^2\left[\mpl^2-12\lambda Y g_5-28\lambda Y^2 g_5'-8\lambda Y^3 g_5''\right]-3H^2e^{2\alpha\lambda\phi}(2g_4-8Y g_4'-8Y^2 g_4'')+2H^3\dot\phi e^{-2\lambda\phi}(5Y g_5'+2Y^2 g_5''), \label{eq: GGG-full-Fr}\\
\mathcal{P}&=2\mathcal{G}_T\dot H+e^{2(\alpha+1)\lambda\phi}p_1+8H\dot\phi\alpha\lambda e^{2\alpha\lambda\phi}(g_4-2Yg_4')+H^2\left[3\mpl^2-4\lambda Y(3g_5+2Yg_5')\right]\notag\\
&\quad +6H^2e^{2\alpha\lambda\phi}(g_4-2Yg_4')-4H^3\dot\phi e^{-2\lambda\phi}Yg_5',
\end{align}
respectively.
Substituting Eqs.~(\ref{ansatz-phi}) and~(\ref{eq: GGG-H}) into Eq.~(\ref{eq: GGG-full-Fr}) yields the leading-order terms of $\cal{E}$ and $\mathcal{P}$ in asymptotic past:
\begin{align}
\mathcal{E}&=e^{2(\alpha+1)\lambda\phi}\biggl[\hat\rho_1(Y_0)+\mathcal{O}(|t|^{-2\alpha})\biggr],\\
\mathcal{P}&=e^{2(\alpha+1)\lambda\phi}\biggl[2e^{-2(\alpha+1)\lambda\phi}\mathcal{G}_1\dot H+\hat p_1(Y_0)+\mathcal{O}(|t|^{-2\alpha})\biggr]. \label{eq: evo-GGG-full}
\end{align}
Note that $\hat\rho_1(Y_0)$ in $\mathcal{E}$ and both the first and second terms of the bracket 
in Eq.~(\ref{eq: evo-GGG-full}) are constant, and thus the $\mathcal{O}(|t|^{-2\alpha})$-terms with $\alpha>0$ are sub-leading. 

In the new framework, the Friedmann and evolution equations are of the form~\cite{Nishi:2016ljg}
\begin{align}
\mathcal{E}&=e^{2(1+\beta)\lambda\phi}\hat\rho_2(Y)+6e^{2\beta\lambda\phi}H\dot\phi Yg'_3+6e^{-2\gamma\lambda\phi}H^2Y(3A'+12YA''+4Y^2A''')\notag\\
&\quad-2e^{-2(1+\beta+2\gamma)\lambda\phi}H^3\dot\phi Y(30B'+75YB''+36Y^2B'''+4Y^3B''''), \label{eq: New-full-Fr}\\
\mathcal{P}&=2\mathcal{G}_T\dot H+e^{2(1+\beta)\lambda\phi}\hat p_2(Y)+8\gamma\lambda H\dot\phi e^{-2\gamma\lambda\phi}Y(A'+2YA'')-4(1+2\beta+4\gamma)\lambda H^2 e^{-(2\beta+2\gamma)\lambda\phi}Y^2\notag\\
&\quad\ \times(6B'+9YB''+2Y^2B''')-6H^2 e^{-2\gamma\lambda\phi}Y(A'+2YA'')+4H^3\dot\phi e^{-2(1+\beta+2\gamma)\lambda\phi}Y(6B'+9YB''+2Y^2B'''),
\end{align}
and the leading-order of which in asymptotic past can be obtained as
\begin{align}
\mathcal{E}&=e^{2(\alpha+1)\lambda\phi}\biggl[\hat\rho_2(Y_0)+\mathcal{O}(|t|^{-2(\beta+\gamma)})\biggr], \label{eq: fr-new-app}\\
\mathcal{P}&=e^{2(\beta+1)\lambda\phi}\biggl[2e^{-2(\beta+1)\lambda\phi}\mathcal{G}_2\dot H+\hat p_2(Y_0)+\mathcal{O}(|t|^{-2(\beta+\gamma)})\biggr]. \label{eq: evo-new-app}
\end{align}
We note that, as long as $\beta+\gamma>0$, the $\mathcal{O}(|t|^{-2(\beta+\gamma)})$-terms of the brackets in Eqs.~(\ref{eq: fr-new-app}) and~(\ref{eq: evo-new-app}) are sub-leading since $\hat\rho_2(Y_0)$ and both the first and second terms of the bracket in Eq.~(\ref{eq: evo-new-app}) are constant.

\section{Quadratic and Cubic Actions}\label{sec: App-pert}
By substituting the perturbed metric, Eqs.~(\ref{eq: pert-met1}) and~(\ref{eq: pert-met2}), into the Horndeski action (\ref{eq: Horndeski action}) and expanding it up to cubic order in the scalar and tensor perturbations, one obtains the perturbed actions,
\begin{align}
S=S^{(2)}+S^{(3)},
\end{align}
where $S^{(2)}$ and $S^{(3)}$ denote the quadratic and cubic actions, respectively. One can eliminate the auxiliary fields $\delta n$ and $\chi$ by solving the constraint equations, and the solutions of which are
\begin{align}
\delta n&=\frac{\mathcal{G}_T}{\Theta}\dot\zeta,\\
\chi&=\frac{1}{a\mathcal{G}_T}\biggl(a^3\mathcal{G}_S\psi-\frac{a\mathcal{G}_T^2}{\Theta}\zeta\biggr),
\end{align}
where $\psi:=\partial^{-2}\dot\zeta$. First, the quadratic action takes the form
\begin{align}
S^{(2)}=S^{(2)}_\zeta+S^{(2)}_h,
\end{align}
where $S^{(2)}_\zeta$ and $S^{(2)}_h$ are given in Eqs.~(\ref{eq: quad-sc}) and~(\ref{eq: quad-tens}), and the coefficients of the quadratic actions read~\cite{Kobayashi:2011nu}
\begin{align}
\mathcal{G}_S&:=\mathcal{G}_T\biggl(\frac{\Sigma\mathcal{G}_T}{\Theta^2}+3\biggr),\\
\mathcal{F}_S&:=\frac{1}{a}\frac{\D}{\D t}\biggl(\frac{a\mathcal{G}_T^2}{\Theta}\biggr)-\mathcal{F}_T,\\
\mathcal{G}_T&:=2\left[G_4-2XG_{4X}-X(H\dot\phi G_{5X}-G_{5\phi})\right],\\
\mathcal{F}_T&:=2\left[G_4-X(\ddot\phi G_{5X}+G_{5\phi})\right],
\end{align}
with
\begin{align}
\Theta&:=-\dot{\phi}XG_{3X}+2HG_4-8HXG_{4X}\notag-8HX^2G_{4XX}+\dot{\phi}G_{4\phi}+2X\dot{\phi}G_{4\phi{X}}\notag\\&
\quad-H^2\dot{\phi}(5XG_{5X}+2X^2G_{5XX})+2HX(3G_{5\phi}+2XG_{5\phi{X}}), \label{Theta-ap}\\
\Sigma&:=XG_{2X}+2X^2G_{2XX}+12H\dot{\phi}XG_{3X}+6H\dot{\phi}X^2G_{3XX}-2XG_{3\phi}-2X^2G_{3\phi{X}}-6H^2G_4\notag\\ &
\quad +6\bigl[H^2(7XG_{4X}+16X^2G_{4XX}+4X^3G_{4XXX})-H\dot{\phi}(G_{4\phi}+5XG_{4\phi{X}}+2X^2G_{4\phi{X}X})\bigr]\notag\\ &
\quad +2H^3\dot{\phi}\left(15XG_{5X}+13X^2G_{5XX}+2X^3G_{5XXX}\right)-6H^2X(6G_{5\phi}+9XG_{5\phi{X}}+2X^2G_{5\phi{X}X}). \label{Sigma-ap}
\end{align}
Note that $\mathcal{G}_1$ in Eq.~(\ref{eq: GGG-evo}) and $\mathcal{G}_2$ in Eq.~(\ref{eq: new-evo}) are corresponding to the leading-order terms of $\mathcal{G}_T$ in each framework. We also note that $\Sigma$ and $\Theta$ can be rewritten as
\begin{align}
\Sigma&=Y\frac{\partial\mathcal{E}}{\partial Y}+\frac{H}{2}\frac{\partial\mathcal{E}}{\partial H}, \label{eq: sigma-compac}\\
\Theta&=-\frac{1}{6}\frac{\partial\cal{E}}{\partial H}. \label{eq: theta-compc}
\end{align}
By using the above and Eqs.~(\ref{eq: GGG-full-Fr}) and~(\ref{eq: New-full-Fr}), these time dependence in Generalized Galilean Genesis and the new framework are obtained, respectively, as
\begin{align}
\Sigma&\propto(-t)^{-2(\alpha+1)}, \label{eq: sigma-ggg}\\
\Theta&\propto(-t)^{-(2\alpha+1)},\label{eq: theta-ggg}
\end{align}
and
\begin{align}
\Sigma&\propto(-t)^{-2(\beta+1)}, \label{eq: sigma-new}\\
\Theta&\propto(-t)^{-(2\beta+1)}. \label{eq: theta-new}
\end{align}
By using the above, one can derive Eqs.~(\ref{eq: gt-ft-time-GGG}), (\ref{eq: gs-fs-time-GGG}), (\ref{eq: gt-time-new}), and~(\ref{eq: ft-time-new}).

The cubic action is of the form
\begin{align}
S^{(3)}=\int\D t\D^3xa^3\biggl[\mathcal{L}_{sss}+\mathcal{L}_{ssh}+\mathcal{L}_{shh}+\mathcal{L}_{hhh}\biggr],
\end{align}
where $\mathcal{L}_{sss},\ \mathcal{L}_{ssh},\  \mathcal{L}_{shh}$, and $\mathcal{L}_{hhh}$ stand for the cubic Lagrangians of the scalar-scalar-scalar, scalar-scalar-tensor, scalar-tensor-tensor, and tensor-tensor-tensor interactions, respectively, and the explicit forms are~\cite{Gao:2011qe,DeFelice:2011uc,Gao:2012ib}
\begin{align}
\mathcal{L}_{sss}&=\biggl[\mathcal{G}_T\biggl(-9\zeta{\dot \zeta}^2+\frac{2\dot{\zeta}}{a^2}\left(\zeta\partial^2\chi+\partial_i\zeta\partial_i\chi\right)+\frac{1}{a^4}\left(\partial_i\chi\right)^2\partial^2\zeta+\frac{1}{2a^4}\zeta\left(\left(\partial^2\chi\right)^2-\left(\partial_i\partial_j\chi\right)^2\right)\biggr)\notag\\
&\ -\mathcal{G}_T\frac{\delta{n}}{a^2}\left(\left(\partial_i\zeta\right)^2+2\zeta\partial^2\zeta\right)+\frac{\mathcal{F}_T}{a^2}\zeta\left(\partial_i\zeta\right)^2+3\Sigma\zeta\delta{n}^2+2\Theta\delta{n}\left(9\zeta\dot{\zeta}-\zeta\partial^2\chi-\partial_i\zeta\partial_i\chi\right)\notag\\
&\ +\mu\left(2{\dot \zeta}^3-\frac{2}{a^2}\partial^2\chi{\dot \zeta}^2+\frac{\dot \zeta}{a^4}\left(\left(\partial^2\chi\right)^2-\left(\partial_i\partial_j\chi\right)^2\right)+4\delta{n}\dot\zeta\frac{\partial^2\zeta}{a^2}-\frac{2\delta{n}}{a^4}\left(\partial^2\zeta\partial^2\chi-\partial_i\partial_j\zeta\partial_i\partial_j\chi\right)\right)\notag\\
&\ +\Gamma\left(3\delta{n}{\dot \zeta}^2-\frac{2}{a^2}\delta{n}\dot{\zeta}\partial^2\chi+\frac{1}{2a^4}\delta{n}\left(\left(\partial^2\chi\right)^2-\left(\partial_i\partial_j\chi\right)^2\right)\right)+\Xi\delta{n}^2\left(\dot{\zeta}-\frac{\partial^2\chi}{3a^2}\right)\notag\\
&\ +\left(\Gamma-\mathcal{G}_T\right)\frac{\delta{n}^2}{a^2}\partial^2\zeta-\frac{1}{3}\left(\Sigma+2X\Sigma_X+H\Xi\right)\delta{n}^3\biggr], \label{eq: cubic-sss-ap}\\
\mathcal{L}_{ssh}&=a\biggl[2\Theta\delta n\partial_i\partial_j\chi h_{ij}+\frac{\Gamma}{2}\delta n\partial_i\partial_j\chi\dot h_{ij}+\frac{\mu}{a^2}\delta n\partial_i\partial_j\chi\partial^2 h_{ij}-\frac{3\mathcal{G}_T}{2}\zeta\partial_i\partial_j\chi\dot h_{ij}-2\mathcal{G}_T\dot\zeta\partial_i\partial_j\chi h_{ij}+\mu\dot\zeta\partial_i\partial_j\chi h_{ij}\notag\\
&\quad\ -\mathcal{F}_T\partial_i\zeta\partial_j\zeta h_{ij}-2\mathcal{G}_T\partial_i\delta n\partial_j\chi h_{ij}+\mu\partial_i\delta n\partial_j \zeta\dot h_{ij}+\frac{\mathcal{G}_T}{2a^2}\partial_k\chi\partial_i\partial_j\chi\partial_k h_{ij}+\frac{\mu}{a^2}\partial_k\chi\partial_i\partial_j\chi\partial_k \dot h_{ij}\biggr],\\
\mathcal{L}_{shh}&=a^3\biggl[\frac{3\mathcal{G}_T}{8}\zeta\dot h_{ij}^2-\frac{\mathcal{F}_T}{8a^2}\zeta(\partial_k h_{ij})^2-\frac{\mu}{4}\dot\zeta\dot h_{ij}^2-\frac{\Gamma}{8}\delta n\dot h_{ij}^2-\frac{\mathcal{G}_T}{8a^2}\delta n(\partial_k h_{ij})^2\notag\\
&\quad -\frac{\mu}{2a^2}\delta n\dot h_{ij}\partial^2 h_{ij}-\frac{\mathcal{G}_T}{4a^2}\partial_k\chi\dot h_{ij}\partial_k h_{ij}-\frac{\mu}{2a^2}\biggl(\partial_i\partial_j\chi\dot h_{ik}\dot h_{jk}-\frac{1}{2}\partial^2\chi\dot h_{ij}^2\biggr)\biggr],\\
\mathcal{L}_{hhh}&=a^3\biggl[\frac{\mathcal{F}_T}{4a^2}\biggl(h_{ik}h_{jl}-\frac{1}{2}h_{ij}h_{kl}\biggr)\partial_k\partial_l h_{ij}+\frac{\mu}{12}\dot h_{ij}^3\biggr],
\end{align}
where
\begin{align}
\Gamma&:=2G_4-8XG_{4X}-8X^2G_{4XX}-2H\dot\phi(5XG_{5X}+2X^2G_{5XX})+2X(3G_{5\phi}+2XG_{5\phi X}), \label{Gamma-ap}\\
\Xi&:=12\dot\phi{X}G_{3X}+6\dot\phi{X^2}G_{3XX}-12HG_4+6\biggl[2H(7XG_{4X}+16X^2G_{4XX}+4X^3G_{4XXX})-\dot\phi(G_{4\phi}\notag\\
&\quad+5XG_{4\phi{X}}+2X^2G_{4\phi{XX}})\biggr]+90H^2\dot\phi{X}G_{5X}+78H^2\dot\phi{X^2}G_{5XX}+12H^2\dot\phi{X^3}G_{5XXX}\notag\\
&\quad-12HX(6G_{5\phi}+9XG_{5\phi X}+2X^2G_{5\phi XX}), \label{Xi-ap}\\
\mu&:=\dot\phi X G_{5X}. \label{eq: def-mu}
\end{align}
Note that $\Xi$ and $\Gamma$ can be written as
\begin{align}
\Xi&=\frac{\partial\Sigma}{\partial H},\\
\Gamma&=\frac{\partial\Theta}{\partial H}.
\end{align}
Then by using Eqs.~(\ref{eq: GGG-full-Fr}),~(\ref{eq: New-full-Fr}),~(\ref{eq: sigma-compac}), and~(\ref{eq: theta-compc}), the time dependence of the above quantities in Generalized Galilean Genesis and the new framework are obtained, respectively, as
\begin{align}
\Xi&\propto(-t)^{-(2\alpha+1)}, \label{eq: Xi-ggg}\\
\Gamma&={\rm const.},\label{eq: Gamma-ggg}
\end{align}
and
\begin{align}
\Xi&\propto(-t)^{-(2\beta+1)}, \label{eq: Xi-new-time}\\
\Gamma&\propto(-t)^{2\gamma}. \label{eq: Gamma-new}
\end{align}
Also, the time dependence of $\mu$ is derived from Eqs.~(\ref{ansatz-phi}) and (\ref{Lagr.new}) as $\mu\propto(-t)^{1+2\beta+4\gamma}$.

By taking into account the time dependence of $\mathcal{G}_S, \mathcal{F}_S, \mathcal{G}_T, \mathcal{F}_T, \Theta, \Sigma, \Xi, \Gamma$, and $\mu$, one can write down the components of the cubic actions as written in Eqs.~(\ref{eq: cubic-s-components-GGG}), (\ref{eq: cano-cubic-s-components-GGG}), and (\ref{eq: cubic-s-components-new}), and also those of the other cubic actions in the new framework used in Sec.~\ref{sec: strong coupling} as
\begin{align}
\mathcal{L}^{(3)}&\supset (-t)^{2(2\beta+3\gamma)}(\partial_t)^2\zeta^2 h_{ij},\ (-t)^{1+4\beta+6\gamma}\partial_t(\partial_i)^2\zeta^2 h_{ij},\ (-t)^{2\gamma}(\partial_i)^2\zeta^2 h_{ij},\ (-t)^{2(1+2\beta+3\gamma)}(\partial_i)^4\zeta^2 h_{ij}\notag\\
&\quad\ (-t)^{2(\beta+2\gamma)}(\partial_t)^2\zeta h_{ij}^2,\ (-t)^{1+2\beta+4\gamma}(\partial_t)(\partial_i)^2\zeta h_{ij}^2,\ (-t)^{2\gamma}(\partial_i)^2\zeta h_{ij}^2,\ (-t)^{2\gamma}(\partial_i)^2 h_{ij}^3, \label{eq: strong-new-cross}
\end{align}
where we imposed $\mu=0$ and $\Gamma=0$ in light of the strong coupling argument.

For the calculation of the scalar non-Gaussianity, it is convenient to derive the following expression~\cite{Gao:2011qe,DeFelice:2011uc,Gao:2012ib}:
\begin{align}
S_\zeta^{(3)}&=\int{{\D}t{\D}^3x}a^3
{\cal G}_S
\biggl\{\frac{\Lambda_1}{H}
\dot\zeta^3+\Lambda_2\zeta\dot\zeta^2+\Lambda_3\zeta\frac{\left(\partial_i\zeta\right)^2}{a^2}
+\frac{\Lambda_4}{H^2}
\dot\zeta^2\frac{\partial^2\zeta}{a^2}
+\Lambda_5\dot\zeta\partial_i\zeta\partial_i
\psi+\Lambda_6\partial^2\zeta\left(\partial_i\psi\right)^2\notag\\
&\quad\quad\quad\quad\quad\quad\quad
+\frac{\Lambda_7}{H^2}
\frac{1}{a^4}\left[\partial^2\zeta\left(\partial_i\zeta\right)^2-\zeta\partial_i\partial_j\left(\partial_i\zeta\partial_j\zeta\right)\right]
+\frac{\Lambda_8}{H}\frac{1}{a^2}\left[\partial^2\zeta\partial_i\zeta\partial_i\psi-\zeta\partial_i\partial_j\left(\partial_i\zeta\partial_j\psi\right)\right]\biggr\}\notag\\
&\quad+\int{{\D}t{\D}^3x}F(\zeta)E_S, \label{cubic}
\end{align}
where 
\begin{align}
\Lambda_{i}&=\sum_{a=0}^1\Lambda_{i,a}\ {\rm for}\ (i\neq3,4),\\
\Lambda_{3}&=\sum_{a=0}^2\Lambda_{3,a},\\
\Lambda_4&:=H^2\biggl[\frac{\Xi}{3}\frac{\mathcal{G}_T^3}{\mathcal{G}_S\Theta^3}+6\frac{\mu\mathcal{G}_T}{\mathcal{G}_S\Theta}+(3\Gamma-\mathcal{G}_T)\frac{\mathcal{G}_T^2}{\mathcal{G}_S\Theta^2}\biggr],
\end{align}
with
\begin{align}
\Lambda_{1,0}&:=H\biggl\{-\frac{H}{3}\frac{\mathcal{G}_T^3\Xi}{\Theta^3\mathcal{G}_S}+\frac{1}{\mathcal{G}_S}\biggl[\frac{3\mathcal{G}_T}{\Theta}(\mathcal{G}_T+\Gamma)+\frac{\Xi\mathcal{G}_T^2}{\Theta^2}+2\mu\biggr]\biggr\},\\
\Lambda_{1,1}&:=H\biggl[\frac{\mathcal{G}_T}{\Theta}\biggl(\frac{1}{c_s^2}-1\biggr)-\frac{\Xi\mathcal{G}_T}{3\Theta^2}-\frac{2\Gamma}{\Theta}-\frac{2\mu}{\mathcal{G}_T}+\frac{2}{3}\frac{\mathcal{G}_T^3}{\Theta^3\mathcal{G}_S}(\Sigma-X\Sigma_X)\biggr],\\
\Lambda_{2,0}&:=3\biggl(1-\frac{H\mathcal{G}_T\mathcal{G}_S}{\Theta\mathcal{F}_S}\biggr),\\
\Lambda_{2,1}&:=\frac{H\mathcal{G}_T\mathcal{G}_S}{\Theta\mathcal{F}_S}(g_T-f_S-f_\Theta),\\
\Lambda_{3,0}&:=\frac{\mathcal{G}_T}{\mathcal{G}_S}\biggl(c_h^2-\frac{H\mathcal{G}_T}{\Theta}\biggr),\\
\Lambda_{3,1}&:=\frac{H\mathcal{G}_T}{\Theta}\biggl[1-\frac{\mathcal{G}_T}{\mathcal{G}_S}(2g_T-f_\Theta)\biggr],\\
\Lambda_{3,2}&:=\frac{H\mathcal{G}_T}{\Theta}(g_T+g_S-f_\Theta),\\
\Lambda_{5,0}&:=-\frac{\mathcal{G}_S}{2\mathcal{G}_T}\biggl(1+\frac{3H\Gamma}{\Theta}+\frac{6\mu H}{\mathcal{G}_T}\biggr),\\
\Lambda_{5,1}&:=-\frac{\mathcal{G}_S}{2\mathcal{G}_T}\biggl[\frac{H\Gamma}{\Theta}(g_T-f_\Gamma+f_\Theta)+\frac{2\mu H}{\mathcal{G}_T}(2g_T-f_\mu)\biggr],\\
\Lambda_{6,0}&:=\frac{3\mathcal{G}_S}{4\mathcal{G}_T}\biggl(1-\frac{H\Gamma}{\Theta}-\frac{2\mu H}{\mathcal{G}_T}\biggr),\\
\Lambda_{6,1}&:=-\frac{\mathcal{G}_S}{4\mathcal{G}_T}\biggl[\frac{H\Gamma}{\Theta}(g_T-f_\Gamma+f_\Theta)+\frac{2\mu H}{\mathcal{G}_T}(2g_T-f_\mu)\biggr],\\
\Lambda_{7,0}&:=\frac{H^2\mathcal{G}_T^3}{6\Theta^2\mathcal{G}_S}\biggl(1-\frac{H\Gamma}{\Theta}-\frac{6\mu H}{\mathcal{G}_T}\biggr),\\
\Lambda_{7,1}&:=\frac{H^2\mathcal{G}_T^3}{\Theta^2\mathcal{G}_S}\biggl[\frac{H\Gamma}{\Theta}\biggl(3g_T-3f_\Theta+f_\Gamma-\frac{3\Theta\mathcal{F}_S}{H\mathcal{G}_T^2}\biggr)+\frac{6\mu H}{\mathcal{G}_T}\biggl(2g_T-2f_\Theta+f_\mu-\frac{2\Theta\mathcal{F}_S}{H\mathcal{G}_T^2}\biggr)\biggr],\\
\Lambda_{8,0}&:=-\frac{H\mathcal{G}_T}{\Theta}\biggl(1-\frac{H\Gamma}{\Theta}-\frac{4\mu H}{\mathcal{G}_T}\biggr),\\
\Lambda_{8,1}&:=-\frac{H\mathcal{G}_T}{2\Theta}\biggl[\frac{H\Gamma}{\Theta}\biggl(g_T+f_\Gamma-2f_\Theta-\frac{2\Theta\mathcal{F}_S}{H\mathcal{G}_T^2}\biggr)-\frac{4\mu H}{\mathcal{G}_T}\biggl(f_\Theta-f_\mu+\frac{\Theta\mathcal{F}_S}{H\mathcal{G}_T^2}\biggr)\biggr].
\end{align}
We also defined
\begin{align}
g_T:=\frac{\dot{\mathcal{G}_T}}{H\mathcal{G}_T},\ g_S:=\frac{\dot{\mathcal{G}_S}}{H\mathcal{G}_S},\ f_S:=\frac{\dot{\mathcal{F}_S}}{H\mathcal{F}_S},\ f_\Theta:=\frac{\dot\Theta}{H\Theta},\ f_\Gamma:=\frac{\dot\Gamma}{H\Gamma},\ f_\mu:=\frac{\dot\mu}{H\mu}.
\end{align}
In the calculation of the non-Gaussianity in Sec.~\ref{sec: non-G}, we used the following time dependence:
\begin{align}
\Lambda_{1,0}, \Lambda_{3,0}, \Lambda_4, \Lambda_{7,0}& \propto(-t)^{-2(\beta+\gamma)},\notag\\ \Lambda_{1,1}, \Lambda_{2,0}, \Lambda_{3,1}, \Lambda_{8,0}&={\rm const.},\notag\\ \Lambda_{1,2}, \Lambda_{3,2}, \Lambda_{5,0}, \Lambda_{6,0}&\propto(-t)^{2(\beta+\gamma)}.
\end{align}
We note that $\Lambda_{5,1},\ \Lambda_{6,1},\ \Lambda_{7,1}$, and $\Lambda_{8,1}$ vanish in the case of $\mu=0=\Gamma$.
\\
\end{widetext}

\bibliography{bibthxGG}

\begin{thebibliography}{67}
\expandafter\ifx\csname natexlab\endcsname\relax\def\natexlab#1{#1}\fi
\expandafter\ifx\csname bibnamefont\endcsname\relax
  \def\bibnamefont#1{#1}\fi
\expandafter\ifx\csname bibfnamefont\endcsname\relax
  \def\bibfnamefont#1{#1}\fi
\expandafter\ifx\csname citenamefont\endcsname\relax
  \def\citenamefont#1{#1}\fi
\expandafter\ifx\csname url\endcsname\relax
  \def\url#1{\texttt{#1}}\fi
\expandafter\ifx\csname urlprefix\endcsname\relax\def\urlprefix{URL }\fi
\providecommand{\bibinfo}[2]{#2}
\providecommand{\eprint}[2][]{\url{#2}}

\bibitem[{\citenamefont{Guth}(1981)}]{Guth:1980zm}
\bibinfo{author}{\bibfnamefont{A.~H.} \bibnamefont{Guth}},
  \bibinfo{journal}{Phys. Rev. D} \textbf{\bibinfo{volume}{23}},
  \bibinfo{pages}{347} (\bibinfo{year}{1981}).

\bibitem[{\citenamefont{Starobinsky}(1980)}]{Starobinsky:1980te}
\bibinfo{author}{\bibfnamefont{A.~A.} \bibnamefont{Starobinsky}},
  \bibinfo{journal}{Phys. Lett. B} \textbf{\bibinfo{volume}{91}},
  \bibinfo{pages}{99} (\bibinfo{year}{1980}).

\bibitem[{\citenamefont{Sato}(1981)}]{Sato:1980yn}
\bibinfo{author}{\bibfnamefont{K.}~\bibnamefont{Sato}}, \bibinfo{journal}{Mon.
  Not. Roy. Astron. Soc.} \textbf{\bibinfo{volume}{195}}, \bibinfo{pages}{467}
  (\bibinfo{year}{1981}).

\bibitem[{\citenamefont{Borde and Vilenkin}(1996)}]{Borde:1996pt}
\bibinfo{author}{\bibfnamefont{A.}~\bibnamefont{Borde}} \bibnamefont{and}
  \bibinfo{author}{\bibfnamefont{A.}~\bibnamefont{Vilenkin}},
  \bibinfo{journal}{Int. J. Mod. Phys. D} \textbf{\bibinfo{volume}{5}},
  \bibinfo{pages}{813} (\bibinfo{year}{1996}), \eprint{gr-qc/9612036}.

\bibitem[{\citenamefont{Creminelli et~al.}(2010)\citenamefont{Creminelli,
  Nicolis, and Trincherini}}]{Creminelli:2010ba}
\bibinfo{author}{\bibfnamefont{P.}~\bibnamefont{Creminelli}},
  \bibinfo{author}{\bibfnamefont{A.}~\bibnamefont{Nicolis}}, \bibnamefont{and}
  \bibinfo{author}{\bibfnamefont{E.}~\bibnamefont{Trincherini}},
  \bibinfo{journal}{JCAP} \textbf{\bibinfo{volume}{11}}, \bibinfo{pages}{021}
  (\bibinfo{year}{2010}), \eprint{1007.0027}.

\bibitem[{\citenamefont{Battefeld and Peter}(2015)}]{Battefeld:2014uga}
\bibinfo{author}{\bibfnamefont{D.}~\bibnamefont{Battefeld}} \bibnamefont{and}
  \bibinfo{author}{\bibfnamefont{P.}~\bibnamefont{Peter}},
  \bibinfo{journal}{Phys. Rept.} \textbf{\bibinfo{volume}{571}},
  \bibinfo{pages}{1} (\bibinfo{year}{2015}), \eprint{1406.2790}.

\bibitem[{\citenamefont{Nishi and Kobayashi}(2015)}]{Nishi:2015pta}
\bibinfo{author}{\bibfnamefont{S.}~\bibnamefont{Nishi}} \bibnamefont{and}
  \bibinfo{author}{\bibfnamefont{T.}~\bibnamefont{Kobayashi}},
  \bibinfo{journal}{JCAP} \textbf{\bibinfo{volume}{03}}, \bibinfo{pages}{057}
  (\bibinfo{year}{2015}), \eprint{1501.02553}.

\bibitem[{\citenamefont{Nishi and Kobayashi}(2017)}]{Nishi:2016ljg}
\bibinfo{author}{\bibfnamefont{S.}~\bibnamefont{Nishi}} \bibnamefont{and}
  \bibinfo{author}{\bibfnamefont{T.}~\bibnamefont{Kobayashi}},
  \bibinfo{journal}{Phys. Rev. D} \textbf{\bibinfo{volume}{95}},
  \bibinfo{pages}{064001} (\bibinfo{year}{2017}), \eprint{1611.01906}.

\bibitem[{\citenamefont{Libanov et~al.}(2016)\citenamefont{Libanov, Mironov,
  and Rubakov}}]{Libanov:2016kfc}
\bibinfo{author}{\bibfnamefont{M.}~\bibnamefont{Libanov}},
  \bibinfo{author}{\bibfnamefont{S.}~\bibnamefont{Mironov}}, \bibnamefont{and}
  \bibinfo{author}{\bibfnamefont{V.}~\bibnamefont{Rubakov}},
  \bibinfo{journal}{JCAP} \textbf{\bibinfo{volume}{08}}, \bibinfo{pages}{037}
  (\bibinfo{year}{2016}), \eprint{1605.05992}.

\bibitem[{\citenamefont{Kobayashi}(2016)}]{Kobayashi:2016xpl}
\bibinfo{author}{\bibfnamefont{T.}~\bibnamefont{Kobayashi}},
  \bibinfo{journal}{Phys. Rev. D} \textbf{\bibinfo{volume}{94}},
  \bibinfo{pages}{043511} (\bibinfo{year}{2016}), \eprint{1606.05831}.

\bibitem[{\citenamefont{Cai et~al.}(2017{\natexlab{a}})\citenamefont{Cai, Wan,
  Li, Qiu, and Piao}}]{Cai:2016thi}
\bibinfo{author}{\bibfnamefont{Y.}~\bibnamefont{Cai}},
  \bibinfo{author}{\bibfnamefont{Y.}~\bibnamefont{Wan}},
  \bibinfo{author}{\bibfnamefont{H.-G.} \bibnamefont{Li}},
  \bibinfo{author}{\bibfnamefont{T.}~\bibnamefont{Qiu}}, \bibnamefont{and}
  \bibinfo{author}{\bibfnamefont{Y.-S.} \bibnamefont{Piao}},
  \bibinfo{journal}{JHEP} \textbf{\bibinfo{volume}{01}}, \bibinfo{pages}{090}
  (\bibinfo{year}{2017}{\natexlab{a}}), \eprint{1610.03400}.

\bibitem[{\citenamefont{Creminelli et~al.}(2016)\citenamefont{Creminelli,
  Pirtskhalava, Santoni, and Trincherini}}]{Creminelli:2016zwa}
\bibinfo{author}{\bibfnamefont{P.}~\bibnamefont{Creminelli}},
  \bibinfo{author}{\bibfnamefont{D.}~\bibnamefont{Pirtskhalava}},
  \bibinfo{author}{\bibfnamefont{L.}~\bibnamefont{Santoni}}, \bibnamefont{and}
  \bibinfo{author}{\bibfnamefont{E.}~\bibnamefont{Trincherini}},
  \bibinfo{journal}{JCAP} \textbf{\bibinfo{volume}{11}}, \bibinfo{pages}{047}
  (\bibinfo{year}{2016}), \eprint{1610.04207}.

\bibitem[{\citenamefont{Akama and Kobayashi}(2017)}]{Akama:2017jsa}
\bibinfo{author}{\bibfnamefont{S.}~\bibnamefont{Akama}} \bibnamefont{and}
  \bibinfo{author}{\bibfnamefont{T.}~\bibnamefont{Kobayashi}},
  \bibinfo{journal}{Phys. Rev. D} \textbf{\bibinfo{volume}{95}},
  \bibinfo{pages}{064011} (\bibinfo{year}{2017}), \eprint{1701.02926}.

\bibitem[{\citenamefont{Cai et~al.}(2017{\natexlab{b}})\citenamefont{Cai, Li,
  Qiu, and Piao}}]{Cai:2017tku}
\bibinfo{author}{\bibfnamefont{Y.}~\bibnamefont{Cai}},
  \bibinfo{author}{\bibfnamefont{H.-G.} \bibnamefont{Li}},
  \bibinfo{author}{\bibfnamefont{T.}~\bibnamefont{Qiu}}, \bibnamefont{and}
  \bibinfo{author}{\bibfnamefont{Y.-S.} \bibnamefont{Piao}},
  \bibinfo{journal}{Eur. Phys. J. C} \textbf{\bibinfo{volume}{77}},
  \bibinfo{pages}{369} (\bibinfo{year}{2017}{\natexlab{b}}),
  \eprint{1701.04330}.

\bibitem[{\citenamefont{Cai and Piao}(2017)}]{Cai:2017dyi}
\bibinfo{author}{\bibfnamefont{Y.}~\bibnamefont{Cai}} \bibnamefont{and}
  \bibinfo{author}{\bibfnamefont{Y.-S.} \bibnamefont{Piao}},
  \bibinfo{journal}{JHEP} \textbf{\bibinfo{volume}{09}}, \bibinfo{pages}{027}
  (\bibinfo{year}{2017}), \eprint{1705.03401}.

\bibitem[{\citenamefont{Kolevatov et~al.}(2017)\citenamefont{Kolevatov,
  Mironov, Sukhov, and Volkova}}]{Kolevatov:2017voe}
\bibinfo{author}{\bibfnamefont{R.}~\bibnamefont{Kolevatov}},
  \bibinfo{author}{\bibfnamefont{S.}~\bibnamefont{Mironov}},
  \bibinfo{author}{\bibfnamefont{N.}~\bibnamefont{Sukhov}}, \bibnamefont{and}
  \bibinfo{author}{\bibfnamefont{V.}~\bibnamefont{Volkova}},
  \bibinfo{journal}{JCAP} \textbf{\bibinfo{volume}{08}}, \bibinfo{pages}{038}
  (\bibinfo{year}{2017}), \eprint{1705.06626}.

\bibitem[{\citenamefont{Boruah et~al.}(2018)\citenamefont{Boruah, Kim, Rouben,
  and Geshnizjani}}]{Boruah:2018pvq}
\bibinfo{author}{\bibfnamefont{S.~S.} \bibnamefont{Boruah}},
  \bibinfo{author}{\bibfnamefont{H.~J.} \bibnamefont{Kim}},
  \bibinfo{author}{\bibfnamefont{M.}~\bibnamefont{Rouben}}, \bibnamefont{and}
  \bibinfo{author}{\bibfnamefont{G.}~\bibnamefont{Geshnizjani}},
  \bibinfo{journal}{JCAP} \textbf{\bibinfo{volume}{08}}, \bibinfo{pages}{031}
  (\bibinfo{year}{2018}), \eprint{1802.06818}.

\bibitem[{\citenamefont{Heisenberg et~al.}(2018)\citenamefont{Heisenberg, Kase,
  and Tsujikawa}}]{Heisenberg:2018wye}
\bibinfo{author}{\bibfnamefont{L.}~\bibnamefont{Heisenberg}},
  \bibinfo{author}{\bibfnamefont{R.}~\bibnamefont{Kase}}, \bibnamefont{and}
  \bibinfo{author}{\bibfnamefont{S.}~\bibnamefont{Tsujikawa}},
  \bibinfo{journal}{Phys. Rev. D} \textbf{\bibinfo{volume}{98}},
  \bibinfo{pages}{123504} (\bibinfo{year}{2018}), \eprint{1807.07202}.

\bibitem[{\citenamefont{Ye and Piao}(2019{\natexlab{a}})}]{Ye:2019frg}
\bibinfo{author}{\bibfnamefont{G.}~\bibnamefont{Ye}} \bibnamefont{and}
  \bibinfo{author}{\bibfnamefont{Y.-S.} \bibnamefont{Piao}},
  \bibinfo{journal}{Commun. Theor. Phys.} \textbf{\bibinfo{volume}{71}},
  \bibinfo{pages}{427} (\bibinfo{year}{2019}{\natexlab{a}}),
  \eprint{1901.02202}.

\bibitem[{\citenamefont{Ye and Piao}(2019{\natexlab{b}})}]{Ye:2019sth}
\bibinfo{author}{\bibfnamefont{G.}~\bibnamefont{Ye}} \bibnamefont{and}
  \bibinfo{author}{\bibfnamefont{Y.-S.} \bibnamefont{Piao}},
  \bibinfo{journal}{Phys. Rev. D} \textbf{\bibinfo{volume}{99}},
  \bibinfo{pages}{084019} (\bibinfo{year}{2019}{\natexlab{b}}),
  \eprint{1901.08283}.

\bibitem[{\citenamefont{Quintin and Yoshida}(2020)}]{Quintin:2019orx}
\bibinfo{author}{\bibfnamefont{J.}~\bibnamefont{Quintin}} \bibnamefont{and}
  \bibinfo{author}{\bibfnamefont{D.}~\bibnamefont{Yoshida}},
  \bibinfo{journal}{JCAP} \textbf{\bibinfo{volume}{02}}, \bibinfo{pages}{016}
  (\bibinfo{year}{2020}), \eprint{1911.06040}.

\bibitem[{\citenamefont{Ageeva et~al.}(2018)\citenamefont{Ageeva, Evseev,
  Melichev, and Rubakov}}]{Ageeva:2018lko}
\bibinfo{author}{\bibfnamefont{Y.~A.} \bibnamefont{Ageeva}},
  \bibinfo{author}{\bibfnamefont{O.~A.} \bibnamefont{Evseev}},
  \bibinfo{author}{\bibfnamefont{O.~I.} \bibnamefont{Melichev}},
  \bibnamefont{and} \bibinfo{author}{\bibfnamefont{V.~A.}
  \bibnamefont{Rubakov}}, \bibinfo{journal}{EPJ Web Conf.}
  \textbf{\bibinfo{volume}{191}}, \bibinfo{pages}{07010}
  (\bibinfo{year}{2018}), \eprint{1810.00465}.

\bibitem[{\citenamefont{Ageeva et~al.}(2020{\natexlab{a}})\citenamefont{Ageeva,
  Evseev, Melichev, and Rubakov}}]{Ageeva:2020gti}
\bibinfo{author}{\bibfnamefont{Y.}~\bibnamefont{Ageeva}},
  \bibinfo{author}{\bibfnamefont{O.}~\bibnamefont{Evseev}},
  \bibinfo{author}{\bibfnamefont{O.}~\bibnamefont{Melichev}}, \bibnamefont{and}
  \bibinfo{author}{\bibfnamefont{V.}~\bibnamefont{Rubakov}},
  \bibinfo{journal}{Phys. Rev. D} \textbf{\bibinfo{volume}{102}},
  \bibinfo{pages}{023519} (\bibinfo{year}{2020}{\natexlab{a}}),
  \eprint{2003.01202}.

\bibitem[{\citenamefont{Ageeva et~al.}(2020{\natexlab{b}})\citenamefont{Ageeva,
  Petrov, and Rubakov}}]{Ageeva:2020buc}
\bibinfo{author}{\bibfnamefont{Y.}~\bibnamefont{Ageeva}},
  \bibinfo{author}{\bibfnamefont{P.}~\bibnamefont{Petrov}}, \bibnamefont{and}
  \bibinfo{author}{\bibfnamefont{V.}~\bibnamefont{Rubakov}},
  \bibinfo{journal}{JHEP} \textbf{\bibinfo{volume}{12}}, \bibinfo{pages}{107}
  (\bibinfo{year}{2020}{\natexlab{b}}), \eprint{2009.05071}.

\bibitem[{\citenamefont{Ageeva et~al.}(2021)\citenamefont{Ageeva, Petrov, and
  Rubakov}}]{Ageeva:2021yik}
\bibinfo{author}{\bibfnamefont{Y.}~\bibnamefont{Ageeva}},
  \bibinfo{author}{\bibfnamefont{P.}~\bibnamefont{Petrov}}, \bibnamefont{and}
  \bibinfo{author}{\bibfnamefont{V.}~\bibnamefont{Rubakov}},
  \bibinfo{journal}{Phys. Rev. D} \textbf{\bibinfo{volume}{104}},
  \bibinfo{pages}{063530} (\bibinfo{year}{2021}), \eprint{2104.13412}.

\bibitem[{\citenamefont{Ageeva and Petrov}(2022)}]{Ageeva:2022fyq}
\bibinfo{author}{\bibfnamefont{Y.}~\bibnamefont{Ageeva}} \bibnamefont{and}
  \bibinfo{author}{\bibfnamefont{P.}~\bibnamefont{Petrov}}
  (\bibinfo{year}{2022}), \eprint{2206.10646}.

\bibitem[{\citenamefont{Ageeva et~al.}(2022)\citenamefont{Ageeva, Petrov, and
  Rubakov}}]{Ageeva:2022asq}
\bibinfo{author}{\bibfnamefont{Y.}~\bibnamefont{Ageeva}},
  \bibinfo{author}{\bibfnamefont{P.}~\bibnamefont{Petrov}}, \bibnamefont{and}
  \bibinfo{author}{\bibfnamefont{V.}~\bibnamefont{Rubakov}}
  (\bibinfo{year}{2022}), \eprint{2207.04071}.

\bibitem[{\citenamefont{Horndeski}(1974)}]{Horndeski:1974wa}
\bibinfo{author}{\bibfnamefont{G.~W.} \bibnamefont{Horndeski}},
  \bibinfo{journal}{Int. J. Theor. Phys.} \textbf{\bibinfo{volume}{10}},
  \bibinfo{pages}{363} (\bibinfo{year}{1974}).

\bibitem[{\citenamefont{Deffayet et~al.}(2011)\citenamefont{Deffayet, Gao,
  Steer, and Zahariade}}]{Deffayet:2011gz}
\bibinfo{author}{\bibfnamefont{C.}~\bibnamefont{Deffayet}},
  \bibinfo{author}{\bibfnamefont{X.}~\bibnamefont{Gao}},
  \bibinfo{author}{\bibfnamefont{D.~A.} \bibnamefont{Steer}}, \bibnamefont{and}
  \bibinfo{author}{\bibfnamefont{G.}~\bibnamefont{Zahariade}},
  \bibinfo{journal}{Phys. Rev. D} \textbf{\bibinfo{volume}{84}},
  \bibinfo{pages}{064039} (\bibinfo{year}{2011}), \eprint{1103.3260}.

\bibitem[{\citenamefont{Kobayashi et~al.}(2011)\citenamefont{Kobayashi,
  Yamaguchi, and Yokoyama}}]{Kobayashi:2011nu}
\bibinfo{author}{\bibfnamefont{T.}~\bibnamefont{Kobayashi}},
  \bibinfo{author}{\bibfnamefont{M.}~\bibnamefont{Yamaguchi}},
  \bibnamefont{and} \bibinfo{author}{\bibfnamefont{J.}~\bibnamefont{Yokoyama}},
  \bibinfo{journal}{Prog. Theor. Phys.} \textbf{\bibinfo{volume}{126}},
  \bibinfo{pages}{511} (\bibinfo{year}{2011}), \eprint{1105.5723}.

\bibitem[{\citenamefont{Kobayashi}(2019)}]{Kobayashi:2019hrl}
\bibinfo{author}{\bibfnamefont{T.}~\bibnamefont{Kobayashi}},
  \bibinfo{journal}{Rept. Prog. Phys.} \textbf{\bibinfo{volume}{82}},
  \bibinfo{pages}{086901} (\bibinfo{year}{2019}), \eprint{1901.07183}.

\bibitem[{\citenamefont{Cai et~al.}(2022)\citenamefont{Cai, Xu, Zhao, and
  Zhou}}]{Cai:2022ori}
\bibinfo{author}{\bibfnamefont{Y.}~\bibnamefont{Cai}},
  \bibinfo{author}{\bibfnamefont{J.}~\bibnamefont{Xu}},
  \bibinfo{author}{\bibfnamefont{S.}~\bibnamefont{Zhao}}, \bibnamefont{and}
  \bibinfo{author}{\bibfnamefont{S.}~\bibnamefont{Zhou}}
  (\bibinfo{year}{2022}), \eprint{2207.11772}.

\bibitem[{\citenamefont{Piao}(2011)}]{Piao:2010bi}
\bibinfo{author}{\bibfnamefont{Y.-S.} \bibnamefont{Piao}},
  \bibinfo{journal}{Phys. Lett. B} \textbf{\bibinfo{volume}{701}},
  \bibinfo{pages}{526} (\bibinfo{year}{2011}), \eprint{1012.2734}.

\bibitem[{\citenamefont{Liu et~al.}(2011)\citenamefont{Liu, Zhang, and
  Piao}}]{Liu:2011ns}
\bibinfo{author}{\bibfnamefont{Z.-G.} \bibnamefont{Liu}},
  \bibinfo{author}{\bibfnamefont{J.}~\bibnamefont{Zhang}}, \bibnamefont{and}
  \bibinfo{author}{\bibfnamefont{Y.-S.} \bibnamefont{Piao}},
  \bibinfo{journal}{Phys. Rev. D} \textbf{\bibinfo{volume}{84}},
  \bibinfo{pages}{063508} (\bibinfo{year}{2011}), \eprint{1105.5713}.

\bibitem[{\citenamefont{Perreault~Levasseur
  et~al.}(2011)\citenamefont{Perreault~Levasseur, Brandenberger, and
  Davis}}]{PerreaultLevasseur:2011wto}
\bibinfo{author}{\bibfnamefont{L.}~\bibnamefont{Perreault~Levasseur}},
  \bibinfo{author}{\bibfnamefont{R.}~\bibnamefont{Brandenberger}},
  \bibnamefont{and} \bibinfo{author}{\bibfnamefont{A.-C.} \bibnamefont{Davis}},
  \bibinfo{journal}{Phys. Rev. D} \textbf{\bibinfo{volume}{84}},
  \bibinfo{pages}{103512} (\bibinfo{year}{2011}), \eprint{1105.5649}.

\bibitem[{\citenamefont{Wang and Brandenberger}(2012)}]{Wang:2012bq}
\bibinfo{author}{\bibfnamefont{Y.}~\bibnamefont{Wang}} \bibnamefont{and}
  \bibinfo{author}{\bibfnamefont{R.}~\bibnamefont{Brandenberger}},
  \bibinfo{journal}{JCAP} \textbf{\bibinfo{volume}{10}}, \bibinfo{pages}{021}
  (\bibinfo{year}{2012}), \eprint{1206.4309}.

\bibitem[{\citenamefont{Creminelli et~al.}(2013)\citenamefont{Creminelli,
  Hinterbichler, Khoury, Nicolis, and Trincherini}}]{Creminelli:2012my}
\bibinfo{author}{\bibfnamefont{P.}~\bibnamefont{Creminelli}},
  \bibinfo{author}{\bibfnamefont{K.}~\bibnamefont{Hinterbichler}},
  \bibinfo{author}{\bibfnamefont{J.}~\bibnamefont{Khoury}},
  \bibinfo{author}{\bibfnamefont{A.}~\bibnamefont{Nicolis}}, \bibnamefont{and}
  \bibinfo{author}{\bibfnamefont{E.}~\bibnamefont{Trincherini}},
  \bibinfo{journal}{JHEP} \textbf{\bibinfo{volume}{02}}, \bibinfo{pages}{006}
  (\bibinfo{year}{2013}), \eprint{1209.3768}.

\bibitem[{\citenamefont{Hinterbichler et~al.}(2012)\citenamefont{Hinterbichler,
  Joyce, Khoury, and Miller}}]{Hinterbichler:2012fr}
\bibinfo{author}{\bibfnamefont{K.}~\bibnamefont{Hinterbichler}},
  \bibinfo{author}{\bibfnamefont{A.}~\bibnamefont{Joyce}},
  \bibinfo{author}{\bibfnamefont{J.}~\bibnamefont{Khoury}}, \bibnamefont{and}
  \bibinfo{author}{\bibfnamefont{G.~E.~J.} \bibnamefont{Miller}},
  \bibinfo{journal}{JCAP} \textbf{\bibinfo{volume}{12}}, \bibinfo{pages}{030}
  (\bibinfo{year}{2012}), \eprint{1209.5742}.

\bibitem[{\citenamefont{Hinterbichler et~al.}(2013)\citenamefont{Hinterbichler,
  Joyce, Khoury, and Miller}}]{Hinterbichler:2012yn}
\bibinfo{author}{\bibfnamefont{K.}~\bibnamefont{Hinterbichler}},
  \bibinfo{author}{\bibfnamefont{A.}~\bibnamefont{Joyce}},
  \bibinfo{author}{\bibfnamefont{J.}~\bibnamefont{Khoury}}, \bibnamefont{and}
  \bibinfo{author}{\bibfnamefont{G.~E.~J.} \bibnamefont{Miller}},
  \bibinfo{journal}{Phys. Rev. Lett.} \textbf{\bibinfo{volume}{110}},
  \bibinfo{pages}{241303} (\bibinfo{year}{2013}), \eprint{1212.3607}.

\bibitem[{\citenamefont{Easson et~al.}(2013)\citenamefont{Easson, Sawicki, and
  Vikman}}]{Easson:2013bda}
\bibinfo{author}{\bibfnamefont{D.~A.} \bibnamefont{Easson}},
  \bibinfo{author}{\bibfnamefont{I.}~\bibnamefont{Sawicki}}, \bibnamefont{and}
  \bibinfo{author}{\bibfnamefont{A.}~\bibnamefont{Vikman}},
  \bibinfo{journal}{JCAP} \textbf{\bibinfo{volume}{07}}, \bibinfo{pages}{014}
  (\bibinfo{year}{2013}), \eprint{1304.3903}.

\bibitem[{\citenamefont{Rubakov}(2013)}]{Rubakov:2013kaa}
\bibinfo{author}{\bibfnamefont{V.~A.} \bibnamefont{Rubakov}},
  \bibinfo{journal}{Phys. Rev. D} \textbf{\bibinfo{volume}{88}},
  \bibinfo{pages}{044015} (\bibinfo{year}{2013}), \eprint{1305.2614}.

\bibitem[{\citenamefont{Elder et~al.}(2014)\citenamefont{Elder, Joyce, and
  Khoury}}]{Elder:2013gya}
\bibinfo{author}{\bibfnamefont{B.}~\bibnamefont{Elder}},
  \bibinfo{author}{\bibfnamefont{A.}~\bibnamefont{Joyce}}, \bibnamefont{and}
  \bibinfo{author}{\bibfnamefont{J.}~\bibnamefont{Khoury}},
  \bibinfo{journal}{Phys. Rev. D} \textbf{\bibinfo{volume}{89}},
  \bibinfo{pages}{044027} (\bibinfo{year}{2014}), \eprint{1311.5889}.

\bibitem[{\citenamefont{Cai and Piao}(2016)}]{Cai:2016gjd}
\bibinfo{author}{\bibfnamefont{Y.}~\bibnamefont{Cai}} \bibnamefont{and}
  \bibinfo{author}{\bibfnamefont{Y.-S.} \bibnamefont{Piao}},
  \bibinfo{journal}{JHEP} \textbf{\bibinfo{volume}{03}}, \bibinfo{pages}{134}
  (\bibinfo{year}{2016}), \eprint{1601.07031}.

\bibitem[{\citenamefont{Kobayashi et~al.}(2015)\citenamefont{Kobayashi,
  Yamaguchi, and Yokoyama}}]{Kobayashi:2015gga}
\bibinfo{author}{\bibfnamefont{T.}~\bibnamefont{Kobayashi}},
  \bibinfo{author}{\bibfnamefont{M.}~\bibnamefont{Yamaguchi}},
  \bibnamefont{and} \bibinfo{author}{\bibfnamefont{J.}~\bibnamefont{Yokoyama}},
  \bibinfo{journal}{JCAP} \textbf{\bibinfo{volume}{07}}, \bibinfo{pages}{017}
  (\bibinfo{year}{2015}), \eprint{1504.05710}.

\bibitem[{\citenamefont{Mironov
  et~al.}(2019{\natexlab{a}})\citenamefont{Mironov, Rubakov, and
  Volkova}}]{Mironov:2019qjt}
\bibinfo{author}{\bibfnamefont{S.}~\bibnamefont{Mironov}},
  \bibinfo{author}{\bibfnamefont{V.}~\bibnamefont{Rubakov}}, \bibnamefont{and}
  \bibinfo{author}{\bibfnamefont{V.}~\bibnamefont{Volkova}},
  \bibinfo{journal}{Phys. Rev. D} \textbf{\bibinfo{volume}{100}},
  \bibinfo{pages}{083521} (\bibinfo{year}{2019}{\natexlab{a}}),
  \eprint{1905.06249}.

\bibitem[{\citenamefont{Mironov
  et~al.}(2019{\natexlab{b}})\citenamefont{Mironov, Rubakov, and
  Volkova}}]{Mironov:2019haz}
\bibinfo{author}{\bibfnamefont{S.}~\bibnamefont{Mironov}},
  \bibinfo{author}{\bibfnamefont{V.}~\bibnamefont{Rubakov}}, \bibnamefont{and}
  \bibinfo{author}{\bibfnamefont{V.}~\bibnamefont{Volkova}}
  (\bibinfo{year}{2019}{\natexlab{b}}), \eprint{1906.12139}.

\bibitem[{\citenamefont{Volkova et~al.}(2019)\citenamefont{Volkova, Mironov,
  and Rubakov}}]{Volkova:2019jlj}
\bibinfo{author}{\bibfnamefont{V.~E.} \bibnamefont{Volkova}},
  \bibinfo{author}{\bibfnamefont{S.~A.} \bibnamefont{Mironov}},
  \bibnamefont{and} \bibinfo{author}{\bibfnamefont{V.~A.}
  \bibnamefont{Rubakov}}, \bibinfo{journal}{J. Exp. Theor. Phys.}
  \textbf{\bibinfo{volume}{129}}, \bibinfo{pages}{553} (\bibinfo{year}{2019}).

\bibitem[{\citenamefont{Ilyas et~al.}(2021)\citenamefont{Ilyas, Zhu, Zheng, and
  Cai}}]{Ilyas:2020zcb}
\bibinfo{author}{\bibfnamefont{A.}~\bibnamefont{Ilyas}},
  \bibinfo{author}{\bibfnamefont{M.}~\bibnamefont{Zhu}},
  \bibinfo{author}{\bibfnamefont{Y.}~\bibnamefont{Zheng}}, \bibnamefont{and}
  \bibinfo{author}{\bibfnamefont{Y.-F.} \bibnamefont{Cai}},
  \bibinfo{journal}{JHEP} \textbf{\bibinfo{volume}{01}}, \bibinfo{pages}{141}
  (\bibinfo{year}{2021}), \eprint{2009.10351}.

\bibitem[{\citenamefont{Zhu and Zheng}(2021)}]{Zhu:2021ggm}
\bibinfo{author}{\bibfnamefont{M.}~\bibnamefont{Zhu}} \bibnamefont{and}
  \bibinfo{author}{\bibfnamefont{Y.}~\bibnamefont{Zheng}}
  (\bibinfo{year}{2021}), \eprint{2109.05277}.

\bibitem[{\citenamefont{Gleyzes et~al.}(2015)\citenamefont{Gleyzes, Langlois,
  Piazza, and Vernizzi}}]{Gleyzes:2014dya}
\bibinfo{author}{\bibfnamefont{J.}~\bibnamefont{Gleyzes}},
  \bibinfo{author}{\bibfnamefont{D.}~\bibnamefont{Langlois}},
  \bibinfo{author}{\bibfnamefont{F.}~\bibnamefont{Piazza}}, \bibnamefont{and}
  \bibinfo{author}{\bibfnamefont{F.}~\bibnamefont{Vernizzi}},
  \bibinfo{journal}{Phys. Rev. Lett.} \textbf{\bibinfo{volume}{114}},
  \bibinfo{pages}{211101} (\bibinfo{year}{2015}), \eprint{1404.6495}.

\bibitem[{\citenamefont{Langlois and Noui}(2016)}]{Langlois:2015skt}
\bibinfo{author}{\bibfnamefont{D.}~\bibnamefont{Langlois}} \bibnamefont{and}
  \bibinfo{author}{\bibfnamefont{K.}~\bibnamefont{Noui}},
  \bibinfo{journal}{JCAP} \textbf{\bibinfo{volume}{07}}, \bibinfo{pages}{016}
  (\bibinfo{year}{2016}), \eprint{1512.06820}.

\bibitem[{\citenamefont{Ben~Achour et~al.}(2016)\citenamefont{Ben~Achour,
  Crisostomi, Koyama, Langlois, Noui, and Tasinato}}]{BenAchour:2016fzp}
\bibinfo{author}{\bibfnamefont{J.}~\bibnamefont{Ben~Achour}},
  \bibinfo{author}{\bibfnamefont{M.}~\bibnamefont{Crisostomi}},
  \bibinfo{author}{\bibfnamefont{K.}~\bibnamefont{Koyama}},
  \bibinfo{author}{\bibfnamefont{D.}~\bibnamefont{Langlois}},
  \bibinfo{author}{\bibfnamefont{K.}~\bibnamefont{Noui}}, \bibnamefont{and}
  \bibinfo{author}{\bibfnamefont{G.}~\bibnamefont{Tasinato}},
  \bibinfo{journal}{JHEP} \textbf{\bibinfo{volume}{12}}, \bibinfo{pages}{100}
  (\bibinfo{year}{2016}), \eprint{1608.08135}.

\bibitem[{\citenamefont{Langlois et~al.}(2017)\citenamefont{Langlois,
  Mancarella, Noui, and Vernizzi}}]{Langlois:2017mxy}
\bibinfo{author}{\bibfnamefont{D.}~\bibnamefont{Langlois}},
  \bibinfo{author}{\bibfnamefont{M.}~\bibnamefont{Mancarella}},
  \bibinfo{author}{\bibfnamefont{K.}~\bibnamefont{Noui}}, \bibnamefont{and}
  \bibinfo{author}{\bibfnamefont{F.}~\bibnamefont{Vernizzi}},
  \bibinfo{journal}{JCAP} \textbf{\bibinfo{volume}{05}}, \bibinfo{pages}{033}
  (\bibinfo{year}{2017}), \eprint{1703.03797}.

\bibitem[{\citenamefont{Akrami et~al.}(2020{\natexlab{a}})}]{Planck:2018jri}
\bibinfo{author}{\bibfnamefont{Y.}~\bibnamefont{Akrami}} \bibnamefont{et~al.}
  (\bibinfo{collaboration}{Planck}), \bibinfo{journal}{Astron. Astrophys.}
  \textbf{\bibinfo{volume}{641}}, \bibinfo{pages}{A10}
  (\bibinfo{year}{2020}{\natexlab{a}}), \eprint{1807.06211}.

\bibitem[{\citenamefont{Leblond and Shandera}(2008)}]{Leblond:2008gg}
\bibinfo{author}{\bibfnamefont{L.}~\bibnamefont{Leblond}} \bibnamefont{and}
  \bibinfo{author}{\bibfnamefont{S.}~\bibnamefont{Shandera}},
  \bibinfo{journal}{JCAP} \textbf{\bibinfo{volume}{08}}, \bibinfo{pages}{007}
  (\bibinfo{year}{2008}), \eprint{0802.2290}.

\bibitem[{\citenamefont{Senatore and Zaldarriaga}(2010)}]{Senatore:2009cf}
\bibinfo{author}{\bibfnamefont{L.}~\bibnamefont{Senatore}} \bibnamefont{and}
  \bibinfo{author}{\bibfnamefont{M.}~\bibnamefont{Zaldarriaga}},
  \bibinfo{journal}{JHEP} \textbf{\bibinfo{volume}{12}}, \bibinfo{pages}{008}
  (\bibinfo{year}{2010}), \eprint{0912.2734}.

\bibitem[{\citenamefont{Akama and Hirano}()}]{Akama:prep}
\bibinfo{author}{\bibfnamefont{S.}~\bibnamefont{Akama}} \bibnamefont{and}
  \bibinfo{author}{\bibfnamefont{S.}~\bibnamefont{Hirano}}, \bibinfo{note}{in
  preparation}.

\bibitem[{\citenamefont{Afshordi
  et~al.}(2007{\natexlab{a}})\citenamefont{Afshordi, Chung, and
  Geshnizjani}}]{Afshordi:2006ad}
\bibinfo{author}{\bibfnamefont{N.}~\bibnamefont{Afshordi}},
  \bibinfo{author}{\bibfnamefont{D.~J.~H.} \bibnamefont{Chung}},
  \bibnamefont{and}
  \bibinfo{author}{\bibfnamefont{G.}~\bibnamefont{Geshnizjani}},
  \bibinfo{journal}{Phys. Rev. D} \textbf{\bibinfo{volume}{75}},
  \bibinfo{pages}{083513} (\bibinfo{year}{2007}{\natexlab{a}}),
  \eprint{hep-th/0609150}.

\bibitem[{\citenamefont{Afshordi
  et~al.}(2007{\natexlab{b}})\citenamefont{Afshordi, Chung, Doran, and
  Geshnizjani}}]{Afshordi:2007yx}
\bibinfo{author}{\bibfnamefont{N.}~\bibnamefont{Afshordi}},
  \bibinfo{author}{\bibfnamefont{D.~J.~H.} \bibnamefont{Chung}},
  \bibinfo{author}{\bibfnamefont{M.}~\bibnamefont{Doran}}, \bibnamefont{and}
  \bibinfo{author}{\bibfnamefont{G.}~\bibnamefont{Geshnizjani}},
  \bibinfo{journal}{Phys. Rev. D} \textbf{\bibinfo{volume}{75}},
  \bibinfo{pages}{123509} (\bibinfo{year}{2007}{\natexlab{b}}),
  \eprint{astro-ph/0702002}.

\bibitem[{\citenamefont{Akama and Kobayashi}(2019)}]{Akama:2018cqv}
\bibinfo{author}{\bibfnamefont{S.}~\bibnamefont{Akama}} \bibnamefont{and}
  \bibinfo{author}{\bibfnamefont{T.}~\bibnamefont{Kobayashi}},
  \bibinfo{journal}{Phys. Rev. D} \textbf{\bibinfo{volume}{99}},
  \bibinfo{pages}{043522} (\bibinfo{year}{2019}), \eprint{1810.01863}.

\bibitem[{\citenamefont{Gao and Steer}(2011)}]{Gao:2011qe}
\bibinfo{author}{\bibfnamefont{X.}~\bibnamefont{Gao}} \bibnamefont{and}
  \bibinfo{author}{\bibfnamefont{D.~A.} \bibnamefont{Steer}},
  \bibinfo{journal}{JCAP} \textbf{\bibinfo{volume}{12}}, \bibinfo{pages}{019}
  (\bibinfo{year}{2011}), \eprint{1107.2642}.

\bibitem[{\citenamefont{De~Felice and Tsujikawa}(2011)}]{DeFelice:2011uc}
\bibinfo{author}{\bibfnamefont{A.}~\bibnamefont{De~Felice}} \bibnamefont{and}
  \bibinfo{author}{\bibfnamefont{S.}~\bibnamefont{Tsujikawa}},
  \bibinfo{journal}{Phys. Rev. D} \textbf{\bibinfo{volume}{84}},
  \bibinfo{pages}{083504} (\bibinfo{year}{2011}), \eprint{1107.3917}.

\bibitem[{\citenamefont{Gao et~al.}(2013)\citenamefont{Gao, Kobayashi,
  Shiraishi, Yamaguchi, Yokoyama, and Yokoyama}}]{Gao:2012ib}
\bibinfo{author}{\bibfnamefont{X.}~\bibnamefont{Gao}},
  \bibinfo{author}{\bibfnamefont{T.}~\bibnamefont{Kobayashi}},
  \bibinfo{author}{\bibfnamefont{M.}~\bibnamefont{Shiraishi}},
  \bibinfo{author}{\bibfnamefont{M.}~\bibnamefont{Yamaguchi}},
  \bibinfo{author}{\bibfnamefont{J.}~\bibnamefont{Yokoyama}}, \bibnamefont{and}
  \bibinfo{author}{\bibfnamefont{S.}~\bibnamefont{Yokoyama}},
  \bibinfo{journal}{PTEP} \textbf{\bibinfo{volume}{2013}},
  \bibinfo{pages}{053E03} (\bibinfo{year}{2013}), \eprint{1207.0588}.

\bibitem[{\citenamefont{Akrami et~al.}(2020{\natexlab{b}})}]{Planck:2019kim}
\bibinfo{author}{\bibfnamefont{Y.}~\bibnamefont{Akrami}} \bibnamefont{et~al.}
  (\bibinfo{collaboration}{Planck}), \bibinfo{journal}{Astron. Astrophys.}
  \textbf{\bibinfo{volume}{641}}, \bibinfo{pages}{A9}
  (\bibinfo{year}{2020}{\natexlab{b}}), \eprint{1905.05697}.

\bibitem[{\citenamefont{Baumann et~al.}(2011)\citenamefont{Baumann, Senatore,
  and Zaldarriaga}}]{Baumann:2011dt}
\bibinfo{author}{\bibfnamefont{D.}~\bibnamefont{Baumann}},
  \bibinfo{author}{\bibfnamefont{L.}~\bibnamefont{Senatore}}, \bibnamefont{and}
  \bibinfo{author}{\bibfnamefont{M.}~\bibnamefont{Zaldarriaga}},
  \bibinfo{journal}{JCAP} \textbf{\bibinfo{volume}{05}}, \bibinfo{pages}{004}
  (\bibinfo{year}{2011}), \eprint{1101.3320}.

\bibitem[{\citenamefont{Joyce and Khoury}(2011)}]{Joyce:2011kh}
\bibinfo{author}{\bibfnamefont{A.}~\bibnamefont{Joyce}} \bibnamefont{and}
  \bibinfo{author}{\bibfnamefont{J.}~\bibnamefont{Khoury}},
  \bibinfo{journal}{Phys. Rev. D} \textbf{\bibinfo{volume}{84}},
  \bibinfo{pages}{083514} (\bibinfo{year}{2011}), \eprint{1107.3550}.

\bibitem[{\citenamefont{Libanov et~al.}(2011)\citenamefont{Libanov, Mironov,
  and Rubakov}}]{Libanov:2011bk}
\bibinfo{author}{\bibfnamefont{M.}~\bibnamefont{Libanov}},
  \bibinfo{author}{\bibfnamefont{S.}~\bibnamefont{Mironov}}, \bibnamefont{and}
  \bibinfo{author}{\bibfnamefont{V.}~\bibnamefont{Rubakov}},
  \bibinfo{journal}{Phys. Rev. D} \textbf{\bibinfo{volume}{84}},
  \bibinfo{pages}{083502} (\bibinfo{year}{2011}), \eprint{1105.6230}.

\end{thebibliography}

\end{document}